\documentclass[12pt]{iopart}
\usepackage{graphicx,psfrag,float}
\usepackage{color}
\usepackage{iopams}
\usepackage{amssymb}
\usepackage{amsfonts}
\usepackage{wrapfig}
\usepackage{epsfig}

\newcommand{\ket}[1]{\left|{#1}\right\rangle}

\newcommand{\eket}[2]{\left| \epsilon_{#1}{#2}\right\rangle}
\newcommand{\ebra}[2]{\left\langle \epsilon_{#1}{#2}\right|}

\begin{document}

\title{Persistent dynamic entanglement from classical motion:
How bio-molecular machines can generate non-trivial quantum states}
\author{Gian Giacomo Guerreschi$^{\dag,\ddag}$, Jianming Cai$^{\dag,\ddag,\sharp}$, Sandu Popescu$^{\S}$ and Hans
J. Briegel$^{\dag,\ddag}$}

\address{$^\dag$Institut f{\"u}r Theoretische Physik,
Universit{\"a}t Innsbruck, Technikerstra{\ss }e 25, A-6020
Innsbruck, Austria}
\address{$^\ddag$Institut f\"ur Quantenoptik und Quanteninformation
der \"Osterreichischen Akademie der Wissenschaften, Innsbruck, Austria}
\address{$^\sharp$ Institute for Theoretical Physics,
University Ulm, Albert-Einstein-Allee 11, D-89069 Ulm, Germany}
\address{$^\S$H.H. Wills Physics Laboratory,
University of Bristol, Tyndall Avenue, Bristol BS8 1TL, U.K.}

\date{\today}


\begin{abstract}
Very recently [\emph{Phys. Rev. E \textbf{82}, 021921 (2010)}] a
simple mechanism was presented by which a molecule subjected to forced
oscillations, out of thermal equilibrium, can maintain quantum entanglement
between two of its quantum degrees of freedom.
Crucially, entanglement can be maintained even in the presence of very
intense noise, so intense that no entanglement is possible when the forced
oscillations cease.
This mechanism may allow for the presence of non-trivial quantum entanglement
in biological systems. Here we significantly enlarge the study of this model.
In particular, we show that the persistent generation of dynamic entanglement
is not restricted to the bosonic heat bath model, but it can also be observed
in other decoherence models, e.g. the spin gas model, and in non-Markovian
scenarios.
We also show how conformational changes can be used by an elementary machine
to generate entanglement even in unfavorable conditions.
In biological systems, similar mechanisms could be exploited by more complex
molecular machines or motors.
\end{abstract}

\maketitle

\tableofcontents

\section{Introduction}

The last couple of years have witnessed the birth and explosive development
of a new research area, namely quantum biology.
For many years the possibility of non-trivial quantum effects in biological
systems was thought to be impossible due to the large amount of noise that
characterize such systems.
During the last few years however, very surprisingly, a host of experimental
and theoretical works have pointed to the possible presence of such effects.
Notable results concern the existence of quantum coherence in photosynthesis
\cite{Jang04,Fleming04,Fleming07,Lee07,Scho09,Eng10,Scho10,Mohseni0805,Rebentrost08,Plenio0807,Olaya,Buchleitner,Eisfeld}
and quantum effects in the so called ``chemical compass'' \cite{Ste89,Schu,Ritz,Hore,GG10,Vedral11}.

Of particular interest is the possible existence of non-trivial entanglement
in biological systems.
The reason is that entanglement has dramatic effects, in particular it leads
to a vastly increased capacity of information processing.
Indeed, entanglement is at the core of quantum computation.
The capacity of quantum systems to perform many other tasks is also significantly
increased by entanglement.
Given these benefits it is very tempting to suspect that biological systems
may make use of entanglement.
On the other hand, the existence of entanglement in biological systems was
thought to be even more unlikely than that of mere quantum coherence because
entanglement is extremely fragile and almost impossible to survive in the
presence of noise.
Very recently, however, it was realized \cite{Hans0806}
that the intuition about the fragility of entanglement was based on
considering the behavior of quantum systems at thermal equilibrium.
Biological systems, however, are not at equilibrium - they are open driven
systems, far from thermal equilibrium.
Such systems are capable of ``error correction'' and may actively maintain
entanglement in presence of noise.
This raises the possibility that non-trivial entanglement may actually be
present in biological systems.

In \cite{Cai10} a simple mechanism was presented by which a molecule
subjected to forced oscillations, out of thermal equilibrium, can maintain
quantum entanglement between two of its quantum degrees of freedom.
Crucially, entanglement can be maintained even in the presence of very intense
noise, so intense that no entanglement is possible when the forced oscillations
cease.
A potential biological system in which such an effect may take place is that
of complex molecules (such as proteins) undergoing forced conformational changes.
At present, no such system has been experimentally observed, for the simple reason
that no experiment has yet been carried out to look for it.
To be able to do that it is imperative to first gain a better theoretical
understanding of the model.
This is the scope of the present paper.

The paper is organized as follows: In section \ref{TDHamiltonian} we present
the abstract model of a moving ``two-spin molecule'' with time-dependent coherent
Hamiltonian; then we investigate the system behavior and the entanglement
dynamics under the influence of two distinct environments:
In section \ref{BosonicBath} we consider a bosonic heat bath representative
of a large number of harmonic oscillators, and study its influence both in the
Markovian and non-Markovian regime.
To relax the working assumption of periodic oscillations, we further introduce
stochasticity in the classical motion of the molecule, and include an additional
source of noise to illustrate the robustness of the entanglement generation.
In section \ref{SpinGas} we extend our study to the spin gas model.
The final section \ref{Conclusion} is devoted to some concluding remarks.

\section{How conformational molecular motion can generate entanglement}
\label{TDHamiltonian}

\begin{figure}[htb]
\begin{center}
\includegraphics[width=8cm]{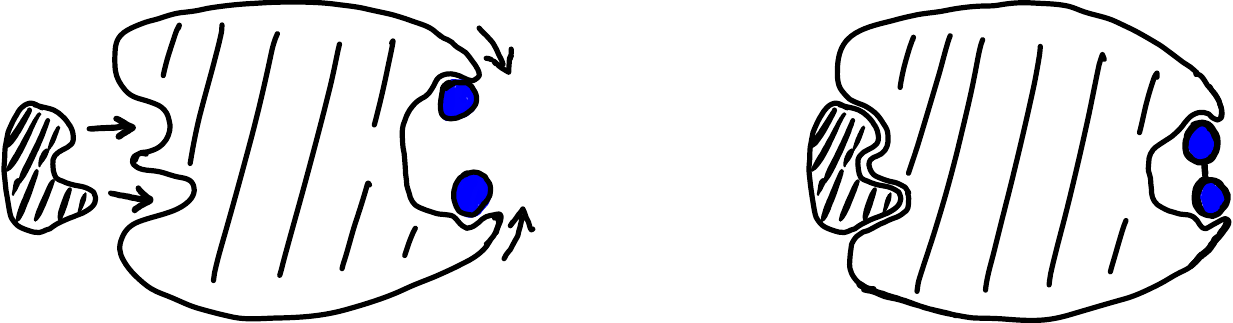}
\end{center}
\caption{(color online) Conformational changes of a bio-molecule \protect\cite{Alberts08},
induced e.g. by the interaction with some other chemical, can lead to a
time-dependent interaction between different sites (blue circles) of the molecule.
See also \protect\cite{Hans0806}.}
\label{AllostericDevice}
\end{figure}

In the following, we consider bio-molecular systems that undergo
controlled conformational changes, using a semi-quantal picture.
We assume that the conformational motion of the molecular structure
can be described, to a very good approximation, classically.
In addition, however, we identify selected quantum degrees of freedom,
embedded in this ``backbone structure'' and localized at several
classically defined sites, which interact with a strength dependent
on their distance (see Fig.~(\ref{AllostericDevice}).

In the following, we consider two quantum degrees of freedom which are forced
to move along some predetermined trajectories and interact via dipolar coupling.
Each of these degrees of freedom shall be represented by a quantum mechanical two-level system.
For easy reference, we call such a model system, a ``two-spin molecule'' \cite{Cai10}.
It can be regarded as an elementary example of a molecular machine which,
as we shall see, is able to persistently generate entanglement.
As such, it also presents a building block which could be exploited in
more complex bio-molecular motors.

\begin{figure}[htb]
\begin{center}
\includegraphics[width=8cm]{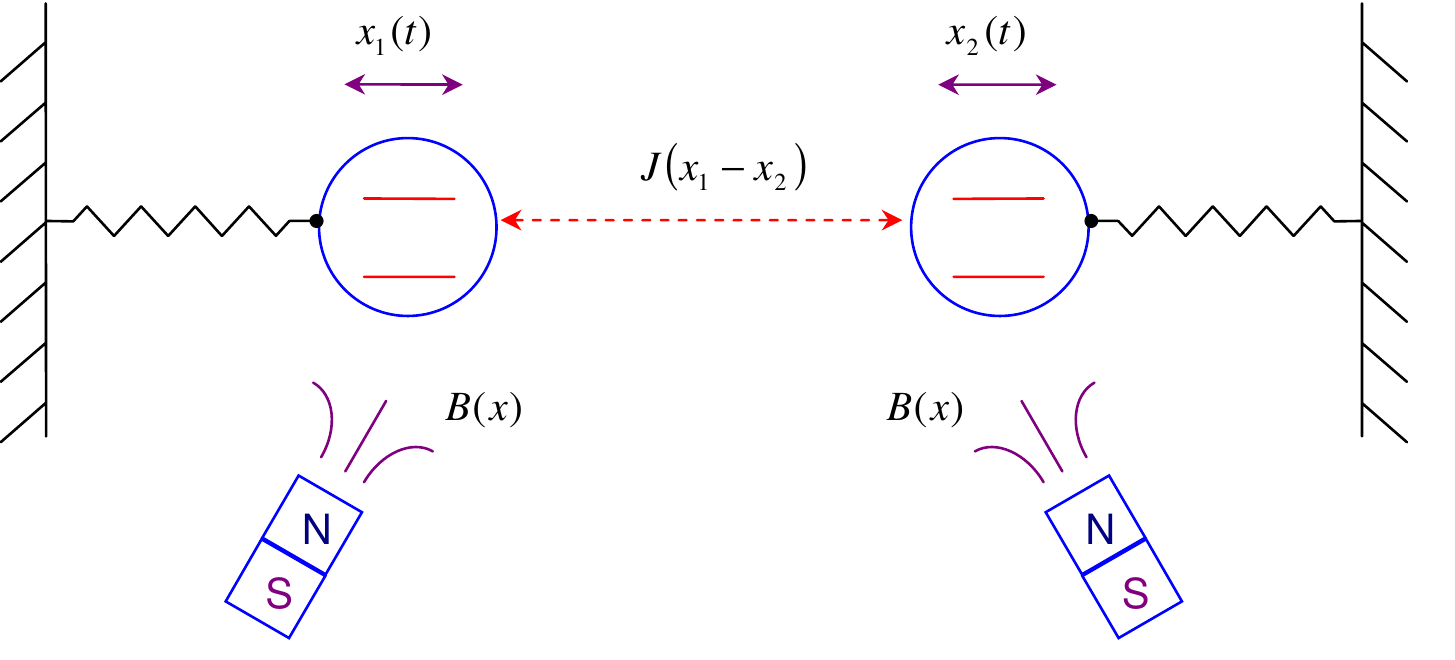}
\end{center}
\caption{A two-spin bio-molecule undergoes conformational changes
such as harmonic oscillation or stochastic motion. Both the
spin-spin interaction strength $J$ and the background field $B$
are position dependent. (Fig. taken from \cite{Cai10})} \label{Conf}
\end{figure}

In more detail, the mechanism is described as follows:
The conformational changes in the molecule force the two localized spins
(or, more generally, two-level systems) to come close or move apart,
inducing an effective time-dependent coupling, both between the two spins
and between each spin and a local field, as described in Fig.~(\ref{Conf}).
Furthermore, it will be assumed (this is essential) that the de-coherent
environment is so noisy that no static entanglement between the two spins
can exist in the static steady state, which is achieved when the molecular
configuration is fixed to any of the possible ones.
As we shall see in the following, if the molecular configuration oscillates
in a periodic way, cyclic generation of fresh entanglement can persist even
in the long time limit.
This feature will persist even in the presence of an additional source of dephasing,
which would generally eliminate all coherence from the system.
Finally, we shall generalize these considerations to more realistic scenarios
involving non-Markovian environments and stochasticity in the classical motion.

In the following, whenever we refer to spins, we may imagine spin-$1/2$ magnetic
moments interacting via a dipolar coupling and subjected to a background magnetic field.
To perform our study, we further assume that the two spins are coupled
with the usual Ising type interaction which depends only on the distance $d$
between the spins, and that there exists a local magnetic field which induces
a (Zeeman) energy level splitting.
The energy splitting is also assumed to depend only on the inter-spin distance $d$, which
thus plays the role of an effective parameter for the overall molecular configuration.
Explicitly, the Hamiltonian for the system is
\begin{equation}
H_{M}(d)=J(d)\sigma _{x}^{(1)}\sigma _{x}^{(2)}+B(d)(\sigma
_{z}^{(1)}+\sigma _{z}^{(2)})  \label{Hamiltonian(d)} ,
\end{equation}
where $\sigma _{x}^{(\alpha )}$, $\sigma _{z}^{(\alpha )}$ are Pauli
operators of the $\alpha$-th spin, $\omega _{0}(d)=2 B(d)$ is the local
level splitting, and we express everything in units of the Planck constant ($\hbar=1$).
Since the molecule undergoes conformational changes, the distance $d$ between
the two spins becomes a function of time: $d\equiv d(t)$.
The explicit time dependence of $H_M(d) \equiv H_M(d(t))$
is determined by the specific choice of the molecular motion:
Initially, we consider forced periodic oscillations, while we later include
an additional stochasticity.
With this model, the molecular motion leads to an effective
time dependent Hamiltonian for the system that we can write as
\begin{equation}
H_M(d(t)) \equiv H_M(t)=J(t)\sigma _{x}^{(1)}\sigma _{x}^{(2)}+B(t)(\sigma
_{z}^{(1)}+\sigma _{z}^{(2)})  \label{Hamiltonian} ,
\end{equation}
where, here and in the following, we use the short notation
$J(t)\equiv J(d(t))$ for all the quantities which depend on
the distance $d$ and so implicitly on the time $t$.
For notational simplicity, we introduce the energy scale
\begin{equation}
\mathcal{E}(t)=\sqrt{4B^{2}(t)+J^{2}(t)}
\end{equation}%
in terms of which the eigenenergies of the system Hamiltonian
$H_M(t)$ and the corresponding eigenstates can be written as
\begin{equation}
\left\{
\begin{array}{l}
\epsilon_0(t)=-\mathcal{E}(t)	\qquad ,  \qquad	\eket{0}{(t)} =(-\eta |00\rangle +|11\rangle )/(1+\eta^{2})^{1/2}\\
\epsilon_1(t)=-J(t)		\qquad ,  \qquad	\eket{1}{(t)} =(|01\rangle -|10\rangle )/\sqrt{2} \\
\epsilon_2(t)=+J(t)		\qquad ,  \qquad	\eket{2}{(t)} =(|01\rangle +|10\rangle )/\sqrt{2} \\
\epsilon_3(t)=+\mathcal{E}(t)	\qquad ,  \qquad	\eket{3}{(t)} =(|00\rangle +\eta |11\rangle)/(1+\eta^{2})^{1/2}
\end{array}
\right. \label{spectra} ,
\end{equation}
with
\begin{equation}
\eta \equiv \eta(t)=\frac{\mathcal{E}(t)-2B(t)}{J(t)} .
\end{equation}%
It can be seen that the energy spectra of the system Hamiltonian
$H_{M}(t)$, and in particular the ground state $|\epsilon _{0}\rangle$,
depends on both the interaction strength $J(t)$ and the local
field $B(t)$, and thus is also time dependent.
In case of weak spin-spin interaction $J(t)\ll B(t)$,
e.g. when two spins are spatially separated from each other, the system
Hamiltonian can be approximated as $H_{M}\sim B(t)(\sigma
_{z}^{(1)}+\sigma _{z}^{(2)})$, and the ground state approaches
the separable state $\ket{11}$. As the spins move closer and
the spin-spin interaction increases, the ground state
changes according to $\eket{0}{(t)}$ in Eq.(\ref{spectra}).
In particular, the entanglement contained in the ground state $\eket{0}{(t)}$
is a monotonically increasing function of the ratio $\frac{J(t)}{B(t)}$,
and can be quantified by the measure of concurrence \cite{Wootters98}.
Concurrence is zero for separable states and reach its maximum value
$C=1$ for maximally entangled states; here it can be calculated as
$C(\eket{0}{(t)}) = | \ebra{0}{(t)}\sigma_{y}\otimes \sigma_{y}\eket{0}{^*(t)} |
        		  = 2\eta /(1+\eta^2)=[1+4(\frac{B(t)}{J(t)})^{2}]^{-1}$,
where $\eket{0}{^*(t)}$ is the complex conjugate of state $\eket{0}{(t)}$
(expressed in the computational product basis, already used in Eq.(\ref{spectra}) ).
In the meantime, the relative energy gap between the ground state and
the first excited state
$(\epsilon_{1}-\epsilon_{0})/(\epsilon_{3}-\epsilon_{0})$
decreases with the ratio $J(t)/B(t)$.

Before taking the environment into account, it is worth looking at
the adiabatic coherent evolution of the oscillating molecule, which
will facilitate our discussions in the following sections. If the
system Hamiltonian changes slowly in time, the instantaneous eigenspaces of
$H_{M}(t)$ will undergo an independent evolution.
The widely used condition that assures the adiabatic dynamics for
closed quantum systems is
\begin{equation}
|\frac{\ebra{k}{(t)}\dot{\epsilon_{l}(t)}\rangle}%
{\epsilon_{k}(t)-\epsilon_{l}(t)}|\ll 1\mbox{,} \quad k \neq l .
\end{equation}
For a detailed discussion concerning the validity of the adiabatic
approximation we refer to \cite{Tong2005,Tong2007}. For the time-dependent system
Hamiltonian introduced in Eq.(\ref{Hamiltonian}) one easily gets
\begin{equation}
|\dot{\epsilon_{1}}(t)\rangle =|\dot{\epsilon_{2}}(t)\rangle=0  \label{DE1}
\end{equation}
and
\begin{equation}
|\dot{\epsilon_{0}}(t)\rangle=-\frac{\dot{\eta}}{1+\eta^{2}}
\eket{3}{(t)}, \quad |\dot{\epsilon_{3}}(t)\rangle=\frac{\dot{\eta}%
}{1+\eta^{2}} \eket{0}{(t)} \label{DE2} .
\end{equation}
Thus the adiabatic condition can be explicitly expressed as
\begin{equation}
|\frac{\dot{\eta}}{1+\eta^{2}}|\ll 2\mathcal{E}(t) .
\label{ACC}
\end{equation}
Whenever the adiabatic condition is satisfied, the molecule that is
initially in an eigenstate of the system Hamiltonian will remain in
the corresponding eigenstate, that is to say
\begin{equation}
\eket{k}{(0)} \rightarrow U(t,0)\eket{k}{(0)}
\simeq e^{-i\int_{0}^{t}\epsilon_{k}(s)d s} \; e^{i\theta_{k}(t)} \eket{k}{(t)} ,
\end{equation}
where $\int_{0}^{t}\epsilon_{k}(s)d s$ is the dynamical phase and
$\theta_{k}(t)=i\int_{0}^{t}\ebra{k}{(s)}\dot{
\epsilon_{k}}(s)\rangle d s$ represents a geometric phase.
One can check from Eq.(\ref{DE1}-\ref{DE2}) that $\ebra{k}{(t)}\dot{\epsilon_{k}}(t)\rangle =0$
is fulfilled $\forall k$ and all times.
Thus no geometric phase is accumulated, i.e. $\theta_{k}(t)=0$ for all $k$,
and the adiabatic evolution can be expressed as follows
\begin{equation}
\eket{k}{(0)} \rightarrow
e^{-i\int_{0}^{t}\epsilon_{k}(s)d s}\eket{k}{(t)} .
\label{ADE}
\end{equation}
%This expression will come up to be useful in the next section.

\section{Bosonic thermal bath}
\label{BosonicBath}

The coherent part of the spin dynamics due to the time-dependent
system Hamiltonian has to be incorporated into a more complex description
which takes into account the presence of an environment.
The actual environment of biological systems is very complex and
probably sophisticated, and far from being thoroughly understood.
However, one can adopt certain representative and well-investigated decoherence
models to gain first insights into the open dynamics of bio-molecules.
We first demonstrate the essential idea by considering the familiar
decoherence model of spins coupled with independent thermal baths
consisting of a large number of harmonic oscillators.
The Hamiltonian of the individual thermal bath coupled to the $\alpha$-th spin ($\alpha=1,2$) is
\begin{equation}
H_{\mathcal{B}}^{(\alpha)}=\sum\limits_{k}\omega_{k}a^{\dagger}_{k,%
\alpha}a_{k,\alpha} ,
\end{equation}
where $a^{\dagger}_{k, \alpha}$ and $a_{k,\alpha}$ are the creation
and annihilation operators for the $k$-th bath oscillator mode and
$\omega_{k}$ the corresponding mode frequency.
The coupling between the spin and the thermal bath is described by
the following interaction Hamiltonian
\begin{equation}
H_{\mathcal{S}\mathcal{B}} = \sum_{\alpha} H_{\mathcal{S}\mathcal{B}}^{(\alpha)}
= \sum_{\alpha}\sigma_{x}^{(\alpha)}
\sum\limits_{k}\nu_{k,\alpha}( a_{k,\alpha}+a^{\dagger}_{k,\alpha}) ,
\label{ISB}
\end{equation}
where $\nu_{k,\alpha}$ is the coupling amplitude to mode $k$.

Within the Born-Markov approximation, and by assuming that the adiabatic
condition for closed quantum systems Eq.(\ref{ACC}) is satisfied,
we have derived a time dependent Lindblad-type quantum master equation
\cite{Cai10} for the system dynamics under the rotating wave approximation,
which reads as
\begin{equation}
\frac{d}{dt}\rho(t)=-i[H_{M}(t),\rho(t)]+\mathcal{D}(t)\rho(t)=\mathcal{L}(t)\rho(t) ,
\label{ME0}
\end{equation}
with the following dissipator
\begin{eqnarray}
\mathcal{D}(t)\rho(t)=\sum_{\alpha,\omega(t)}\Gamma_{\omega(t)}
\left\{A_{\alpha}[\omega(t)]\rho(t)
A^{\dagger}_{\alpha}[\omega(t)]-A^{\dagger}_{\alpha}[\omega(t)]
A_{\alpha}[\omega(t)]\rho(t)\right\}+h.c.  \nonumber \\
A_{\alpha}[\omega(t)]=\sum_{\Delta_{i j}(t)=\omega(t)}
%S_{ij}^{(\alpha)}(t )
\ebra{i}{(t)} \sigma_x^{(\alpha)} \eket{j}{(t)} \times
\eket{i}{(t)} \ebra{j}{(t)} .
\end{eqnarray}
In other words, the time dependence of the system Hamiltonian
introduces an implicit time dependence into the Lindblad generators
$A_{\alpha}[\omega(t)]$, as well as into the rates $\Gamma_{\omega(t)}$
(a sort of one-side Fourier transform of the thermal bath correlation
function, see \cite{Cai10}).
Notice that, as already discussed for Eq.(\ref{Hamiltonian}), the
time dependence of $H_M$ (and so also $\mathcal{D}$ and $\mathcal{L}$)
is actually due to its dependence on the distance $d$ between the spins,
from which the short notation, e.g., $\mathcal{L}(t) \equiv \mathcal{L}(d(t))$.
% \begin{equation}
% L_{\alpha}(\omega(t))=\Gamma^{1/2}_{\omega(t)}\sum_{\Delta_{i
% j}(t)=\omega(t)} S^{(\alpha)}_{ij}(t) \eket{i}{(t)} \ebra{j}{(t)}
% \end{equation}

The bosonic thermal bath is characterized by its spectral density function
$ J(\omega)$ which contains information on both the density of modes in the
frequency domain and the strength of their coupling to the system.
In the following, we choose an Ohmic spectral density
$J(\omega)=\frac{\xi}{2\pi}\omega e^{-\omega/\omega_{c}}$,
with cutoff frequency $\omega_{c}$.
The memory time of the thermal bath strongly depends on the cutoff
frequency and so does its Markovian or non-Markovian behavior \cite{BreuerBook}.
In particular, if $\omega_{c}\gg \omega$, where $\omega$ is
the energy scale of the system transition frequencies, the
reservoir loses its memory quickly, and is called Markovian.
The corresponding correlation function exhibits a sharply peaked kernel
in the limit $\omega_{c}\rightarrow\infty$.
In this case, one can simplify the spectral density as
$J(\omega)=\kappa \omega$ with $\kappa=\frac{\xi}{2\pi}$.
We thus can write the Fourier transformation of
the thermal bath correlation function as $
\Gamma_{\omega(t)}=\kappa\omega(t)(1+ N_{\omega(t),\beta})$
where $N_{\omega(t),\beta}=1/(e^{\omega(t)\beta}-1)$ represents the bosonic
distribution function at inverse temperature $\beta$.

\begin{figure}[htb]
\begin{center}
\includegraphics[width=8cm]{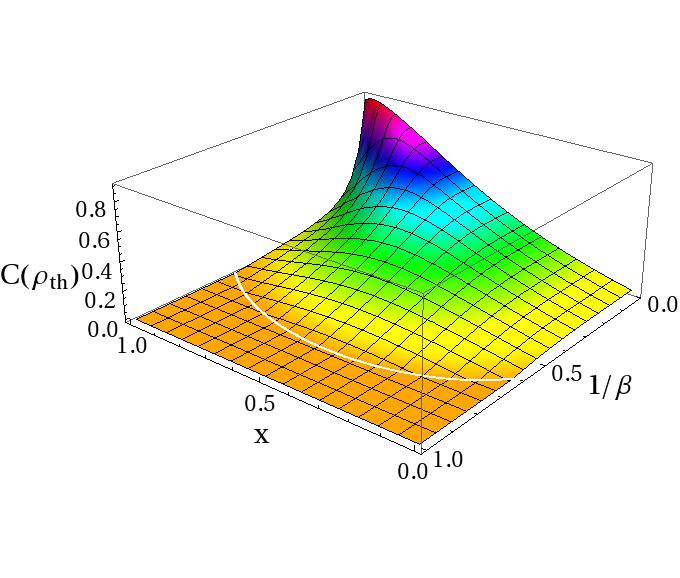}
\end{center}
\vspace{-1.2cm}
\caption{Static entanglement in the thermal
equilibrium state at different inverse temperatures $\beta$ for
$B(x)=B(0)-x$ and $J(x)=J(0)+ x $ with $B(0)=1.2$ and $J(0)=0.1$.
Here $\beta,B,J$ can be varied independently, but it is their relative
value which determines the dynamical behavior.
Therefore, here and in the following, we use dimensionless quantities
obtained by setting the thermal energy $k_{b}T$
as the unit of the energy scale (so that $\beta=1$)
with respect to which all the other parameters are determined,
for example $B=1$ indicates that $\hbar B /k_{b}T=1$.}
\label{FTE}
\end{figure}

In order to establish a comparison between the static and dynamic cases,
we first consider the situation in which the two-spin bio-molecule is
fixed in a static configuration. Since the two spins are coupled to independent
thermal baths at the same temperature, the corresponding quantum master
equation mixes the energy eigenstates in the sense that the molecule
is driven towards its thermal equilibrium state \cite {BreuerBook}
at the reservoir temperature $1/\beta$.
For each spin-spin interaction strength $J(d)=J$ and local
magnetic fields $B(d)=B$, the entanglement of the corresponding
thermal equilibrium state, defined by $\mathcal{L}(d)\rho_{th}=0$,
is quantified by \cite{Cai10}
\begin{equation}
C(\rho_{th})=\frac{2}{\mathcal{Z}}\max {\{0,\frac{2\eta }{1+\eta ^{2}}\sinh
{(\mathcal{E}\beta )}-\cosh {(J\beta )}\}} ,
\end{equation}
with $\mathcal{Z}=2 \cosh{(\mathcal{E} \beta)} + 2 \cosh {(J \beta)}$ the partition function.
After some calculations, it can be seen that the first derivative of $C$
with respect to $\beta$ is always non negative
\begin{equation}
\partial C(\rho_{th})/\partial \beta \geq 0 ,
\end{equation}
which means that the static thermal entanglement decreases as
the temperature increases; in particular, static entanglement
can not survive if the temperature is higher
than a certain critical temperature $1/\beta_{c}$.
In other words, if the molecule stays in a fixed configuration,
entanglement will finally disappear for high enough temperatures,
e.g. see Fig.~(\ref{FTE}).
In this sense, the bosonic heat bath, even though it may not be a very realistic
model of a biological environment, plays a useful role in our investigation.
It is an example of an unfavorable environment, which cannot introduce any
entanglement by itself.

Here we address the key question:
\textit{What happens if one takes into account the molecular motion?}
and in particular: \textit{Can entanglement
possibly build-up through the classical motions of the molecule?}
The answer is affirmative and we demonstrate in the following that
entanglement can indeed persistently recur in an oscillating molecule,
even if the environment is so hot that any static thermal state is separable,
i.e. the temperature $T> \max\{T_{c}\}$ for all possible
molecular configurations encountered during the motion.

\subsection{Qualitative physical picture}
Before proceeding, it is useful to present a simple qualitative
picture which illustrates the idea of how entanglement can be
established through classical motion even in noisy environments.
We consider the following simple process:
at time $t=0$, two spins are far way from each other and
the spin-spin interaction is very weak compared to the local
magnetic fields. Thus the system Hamiltonian can be approximated as
$H_M(0)\simeq B(0)(\sigma_{z}^{(1)}+\sigma_{z}^{(2)})$.
Due to the presence of the thermal bath, the molecule is driven into its thermal
equilibrium state with the corresponding ground state population denoted as $p_{0}$.
If the local energy level splitting $\omega_{0}(0)=2 B(0)$ is comparable or larger
than the thermal energy scale, $\hbar \omega_{0}(0)\geq k_{B}T$, the ground
state population $p_{0}$ at this distant configuration will be relatively
larger than the corresponding populations of the other energy levels.
Nevertheless entanglement does not appear since the ground
state is separable ($\eket{0}{(0)}\simeq\ket{11}$).
Let us further assume that the spins come closer at such a speed that
the coherent evolution due to the system Hamiltonian is adiabatic.
While the spins approach each other, their interaction becomes
stronger and the energy eigenstates adiabatically change into the ones of $H_M(t)$.
In particular the ground state $\eket{0}{(0)} \rightarrow \eket{0}{(t)}$
becomes more and more entangled as the coupling strength between spins increases,
as indicated by the fact that its concurrence $C(\eket{0}{(t)})=J/\mathcal{E}$
is an increasing function of the ratio $J(t)/B(t)$.
Following this reasoning, it is possible that entanglement appears as a transient phenomenon
as long as the adiabatic change of $H_M$ is still fast compared to the thermalization time scale
$\gamma^{-1}$ with $\gamma=\kappa \omega (2 N_{\omega,\beta}+1)$ for
the chosen simple Ohmic spectral density.
\begin{figure}[htb]
\begin{center}
\begin{minipage}{15cm}
\includegraphics[width=6.8cm]{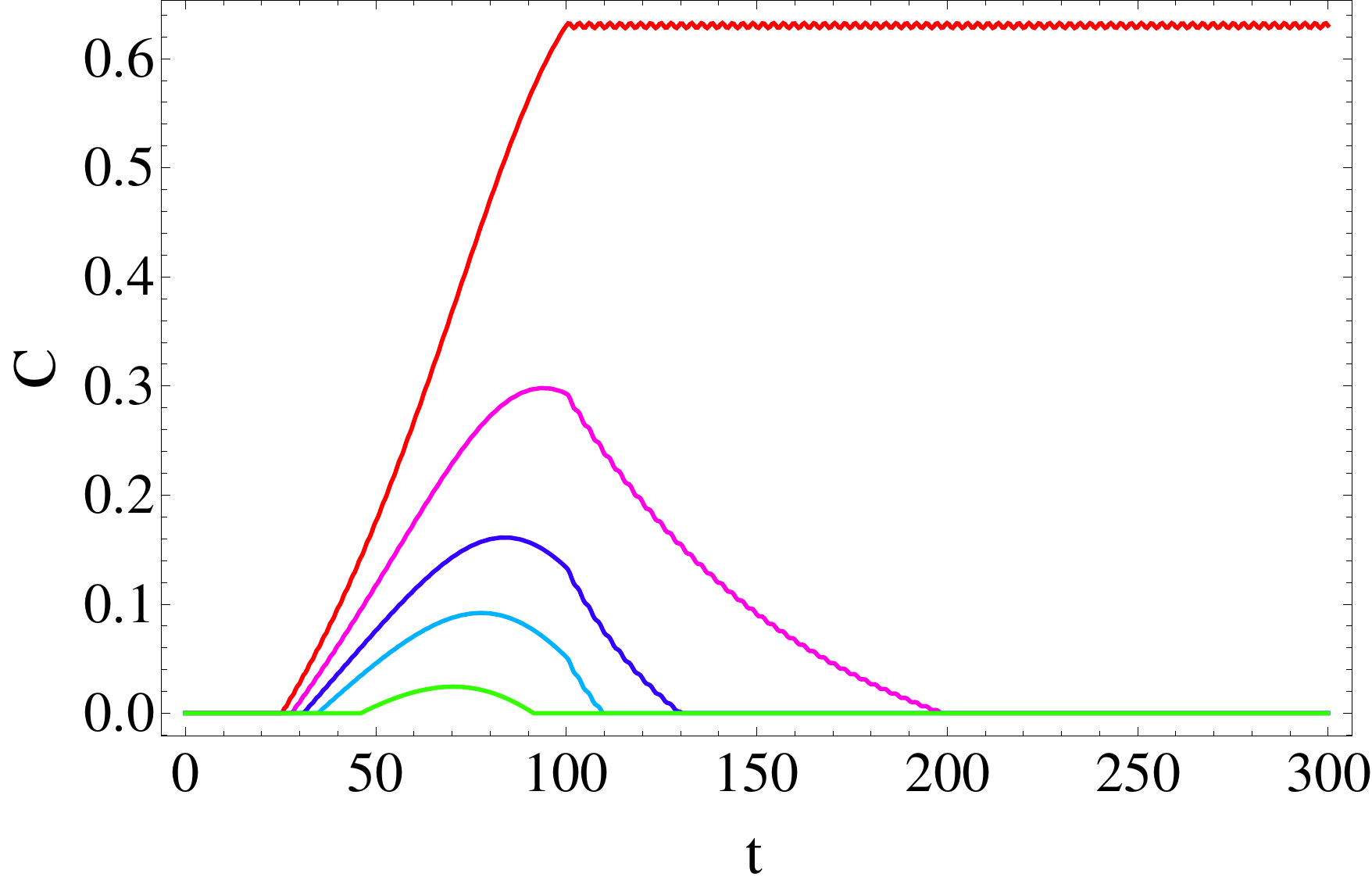}
\hspace{0.5cm}
\includegraphics[width=7cm]{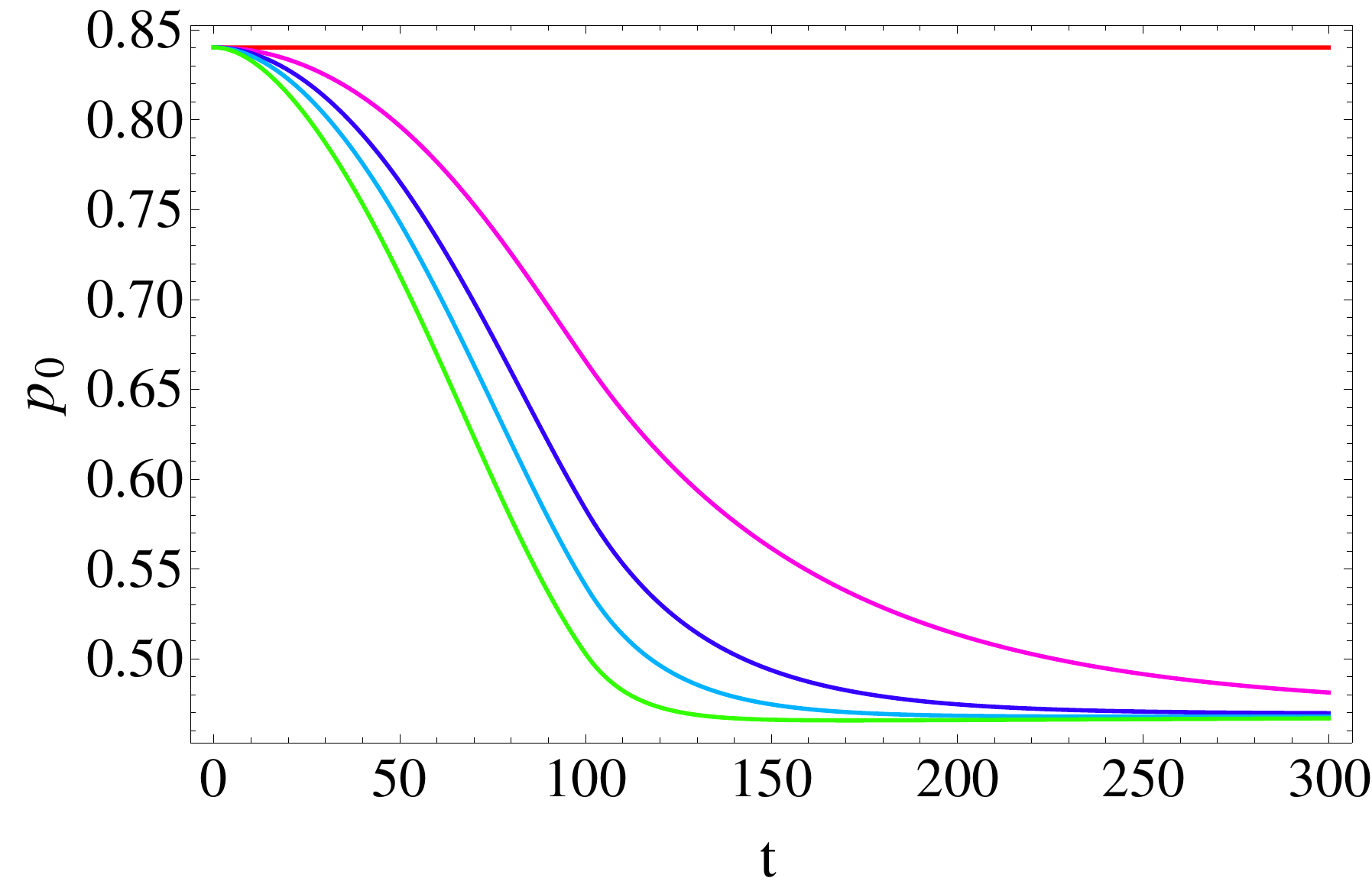}
\end{minipage}
\end{center}
\caption{Entanglement $C$ (left) and the ground state population
$p_{0}$ (right) \textit{vs.} $t$ for a single conformational change with $B(t)$
and $J(t)$ as in Eq.(\protect\ref{DB}-\protect\ref{DJ}) and $B(0)=1.2$, $J(0)=0.1$,
$\protect\delta=0.01$, $t_{0}=100$, with different system-bath coupling strength
(from top to bottom: $\kappa=0$, $0.002$, $0.004$, $0.006$, $0.01$).
Here and in the following, the reservoir inverse temperature is $\protect\beta=1$.}
\label{SinCyl}
\end{figure}

To make the argument more concrete, in Fig.~(\ref{SinCyl}) we illustrate
the above qualitative picture through a simple process, in which
\begin{equation}
B(t)=\left\{
\begin{array}{ll}
B(0)-\delta \cdot t \quad \quad 0\leq t \leq t_{0} &  \\
B(t_{0}) \quad \quad \quad \quad \, t_{0}\leq t &
\end{array}%
\right. ,
\label{DB}
\end{equation}
\begin{equation}
J(t)=\left\{
\begin{array}{ll}
J(0)+\delta \cdot t \quad \quad 0\leq t \leq t_{0} &  \\
J(t_{0}) \quad \quad \quad \quad \, t_{0}\leq t &
\end{array}%
\right. .
\label{DJ}
\end{equation}
Considering a reservoir with inverse temperature $\beta=1$, the initial
ground state population $p_{0}$ is about $85\%$. The change of the
system Hamiltonian is characterized by the small step value
$\delta=0.01$ which ensures the adiabatic condition for closed systems.
It can be seen that \textit{the molecule can be kicked out of its separable
thermal equilibrium state through classical motion} and thus
entanglement grows before being washed out by thermalization.
The dissipative environment now plays its usual destructive role.
Moreover, if the thermalization rate is fast compared to the adiabatic change of
the system Hamiltonian, entanglement is greatly suppressed.

\subsection{Persistent recurrence of fresh entanglement}
\label{Persistent recurrence of fresh entanglement}
To counter-act the detrimental effect of the environment and to sustain
the generation of entanglement as a non-transient phenomenon, a persistent supply
of free energy is needed.
This is an organized form of energy which, as in our mechanism, can be introduced
by conformational changes.
To illustrate our idea, we first consider a two-spin molecule subjected to a deterministic
and periodic oscillatory motion (the strict periodicity will be relaxed in section \ref{SCM}).
The position of the two spins are given by
\begin{equation}
x_{\alpha}(t) =x_{\alpha}(0)+(-1)^{\alpha}a(\cos\frac{2\pi t}{\tau}-1) \qquad \alpha=1,2
\end{equation}
where $x_{\alpha}(0)$ are the initial positions, $a$ is the amplitude
of oscillation, and $\tau$ is the oscillation period.
As a remark, we refer to the next subsection \ref{SCM}
for more general forms of classical motion involving stochasticity.
The coupling between two spins is described by a dipole-dipole Ising
interaction
\begin{equation}
J(t)=J_{0}/d^{3}(t) \quad \mbox{with} \quad d(t)=|x_{1}(t)-x_{2}(t)| .
\end{equation}
while the local magnetic field profile, dependent on the spin position, is chosen as
\begin{equation}
B_{\alpha}(t)=B_0-B_1 e^{-x_{\alpha}^2(t)/\sigma }=B_0-B_1 e^{-d^2(t)/4\sigma } ,
\end{equation}
The choice of $x_{\alpha}(t), B_{\alpha}(t)$ and $J(t)$ is quite arbitrary.
As is explained in the following analysis, the essential condition
for the proper function of the mechanism is that the dominant contribution
to $H_M(t)$ is due to the background field when the spins are far away, and
due to their interaction when they are closest.
This is a natural assumption, since the interaction generally increases when
the distance is reduced.

\begin{figure}[htb]
\begin{center}
\begin{minipage}{15cm}
\includegraphics[width=7cm]{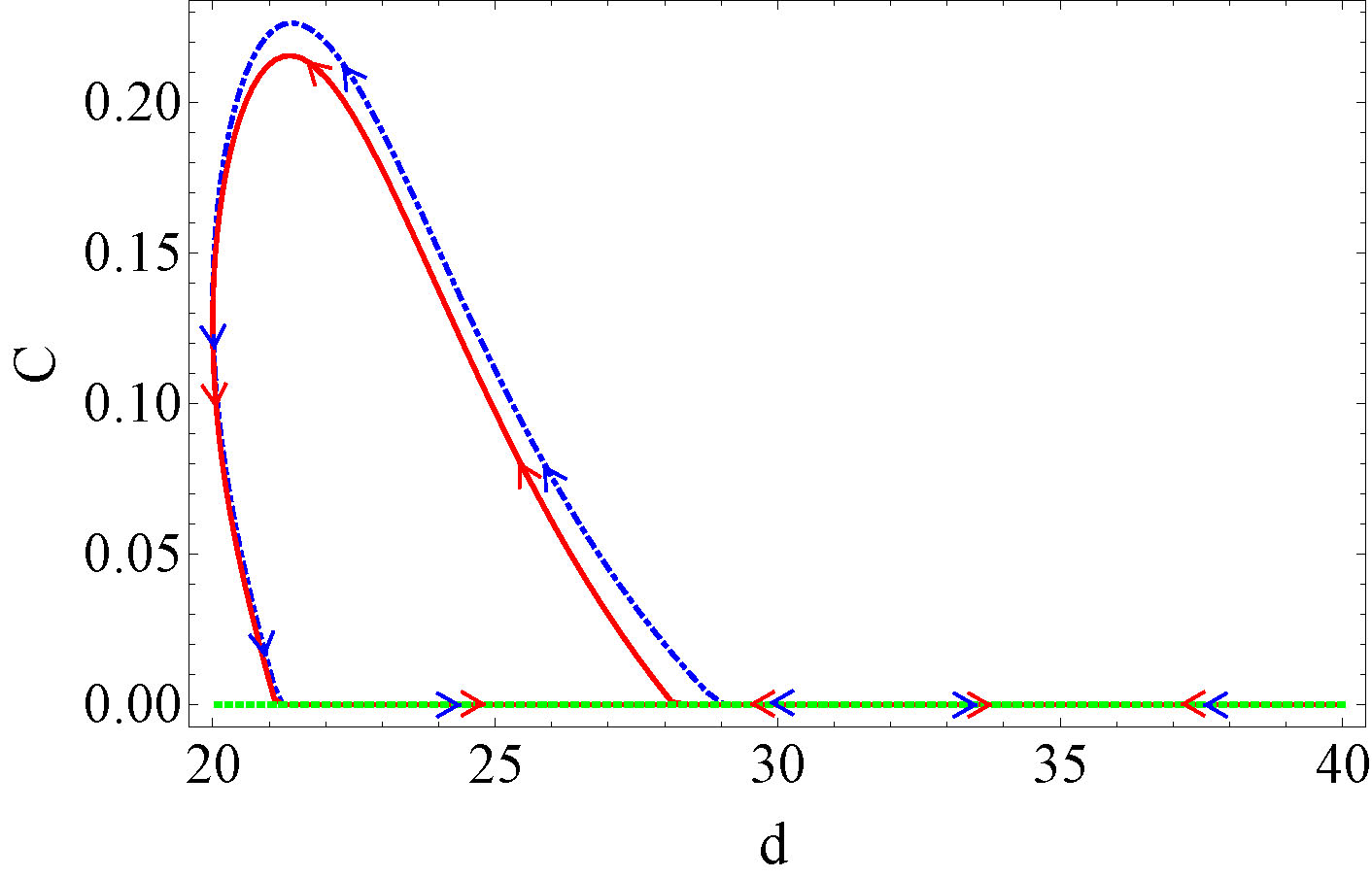}
\hspace{0.5cm}
\includegraphics[width=7cm]{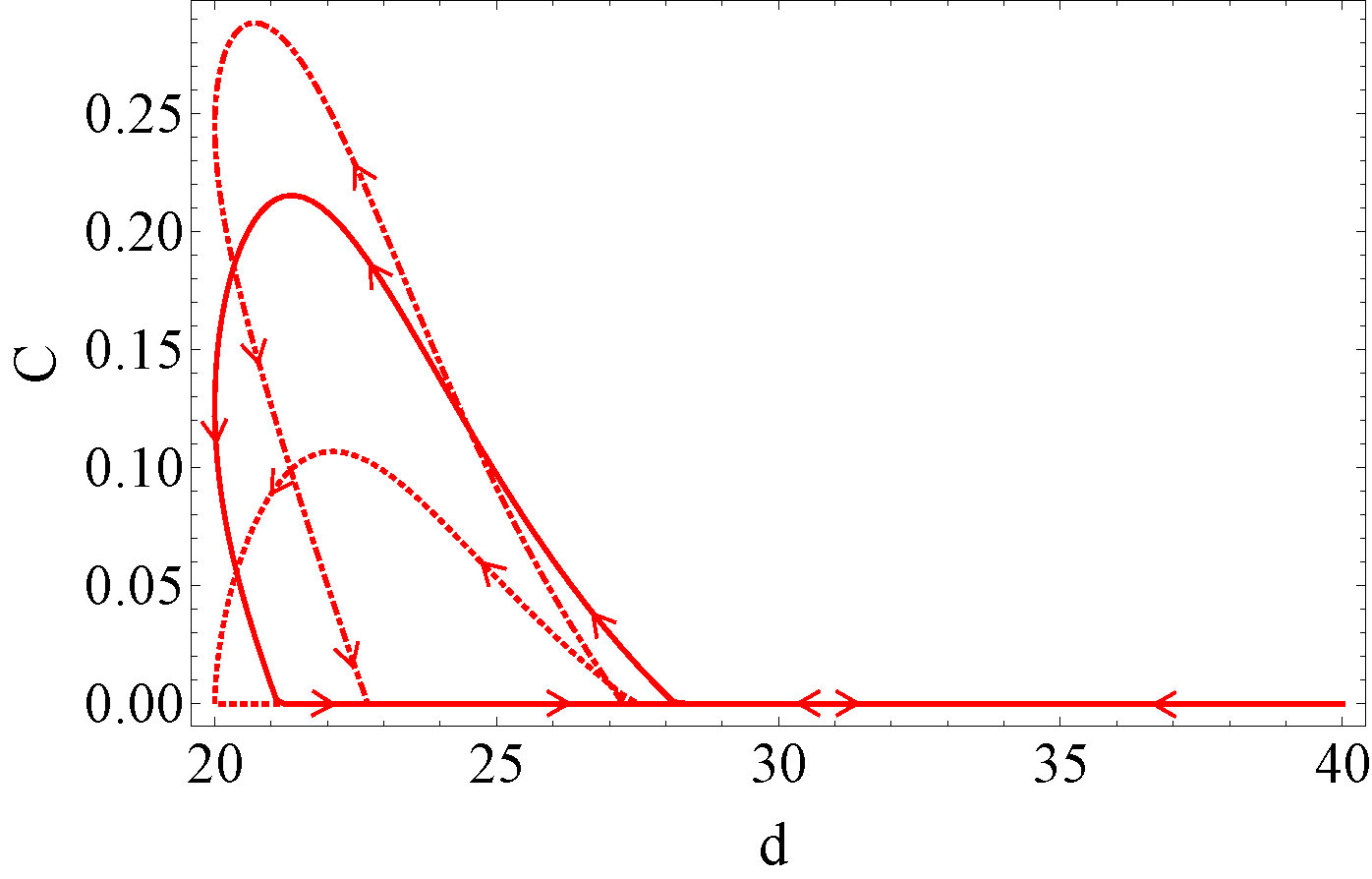}
\end{minipage}
\end{center}
\caption{Left: Entanglement $C$ \textit{vs.} the molecular
configuration $d$ (the distance between the two spins) for the bosonic thermal bath
with $\beta=1$ and $\kappa =0.01$. The dot-dashed, blue curve indicates the first
period; the solid, red curve marks the asymptotic cycle; while the green, dotted
curve refers to the thermal equilibrium case.
The arrows along the curves indicate how the entanglement evolves during the classical
oscillations.
The oscillation parameters are $x_{1}(0)=-x_{2}(0)=-20$, $a=5$, $\tau=100$,
and $B_{0}=1.3$, $B_{1}=2.4$, $\protect\sigma=120$, $J_{0}=10^{4}$.
Right: Entanglement during the asymptotic cycle for different system-bath
coupling strength (from top to bottom $\kappa = 0.005,0.01,0.02$).
% : $\kappa =0.005$, $\kappa =0.01$, $\kappa =0.02$
%Here we assume that the Planck and Boltzmann constant $\hbar=k_{b}=1$.
}\label{QSMEb}
\end{figure}

We assume that the two spins start from the distant configuration and are initially
in the corresponding thermal equilibrium state.
Notice that this particular initial condition is irrelevant for the system asymptotic
dynamics that we are interested in. At time $t=0$ the spins start to oscillate,
\textit{i.e.} they periodically come close to each other and separate again.
The state of the molecule is driven out of thermal equilibrium,
and entanglement is generated.
When the spins move back to their distant configuration, the thermalization
tends to increase the population of the ground state (due to the increasing
energy gap between the ground state and the first excited state), and thus effectively
resets the system to a favorable condition.
Therefore, although environmental noise generally destroys entanglement, it now represents
a crucial factor to enable the persistent and cyclic generation of fresh entanglement.
For simplicity, we call the cycle during the time interval
$[n \tau,(n+1)\tau]$ with $n\rightarrow \infty$ as the asymptotic cycle.
It can be seen from Fig.~(\ref{QSMEb}) that entanglement indeed persistently
recurs during the oscillations even if the thermal bath is so hot that no static entanglement can survive.

\begin{figure}[bht]
\begin{center}
\begin{minipage}{15cm}
\includegraphics[width=6.8cm]{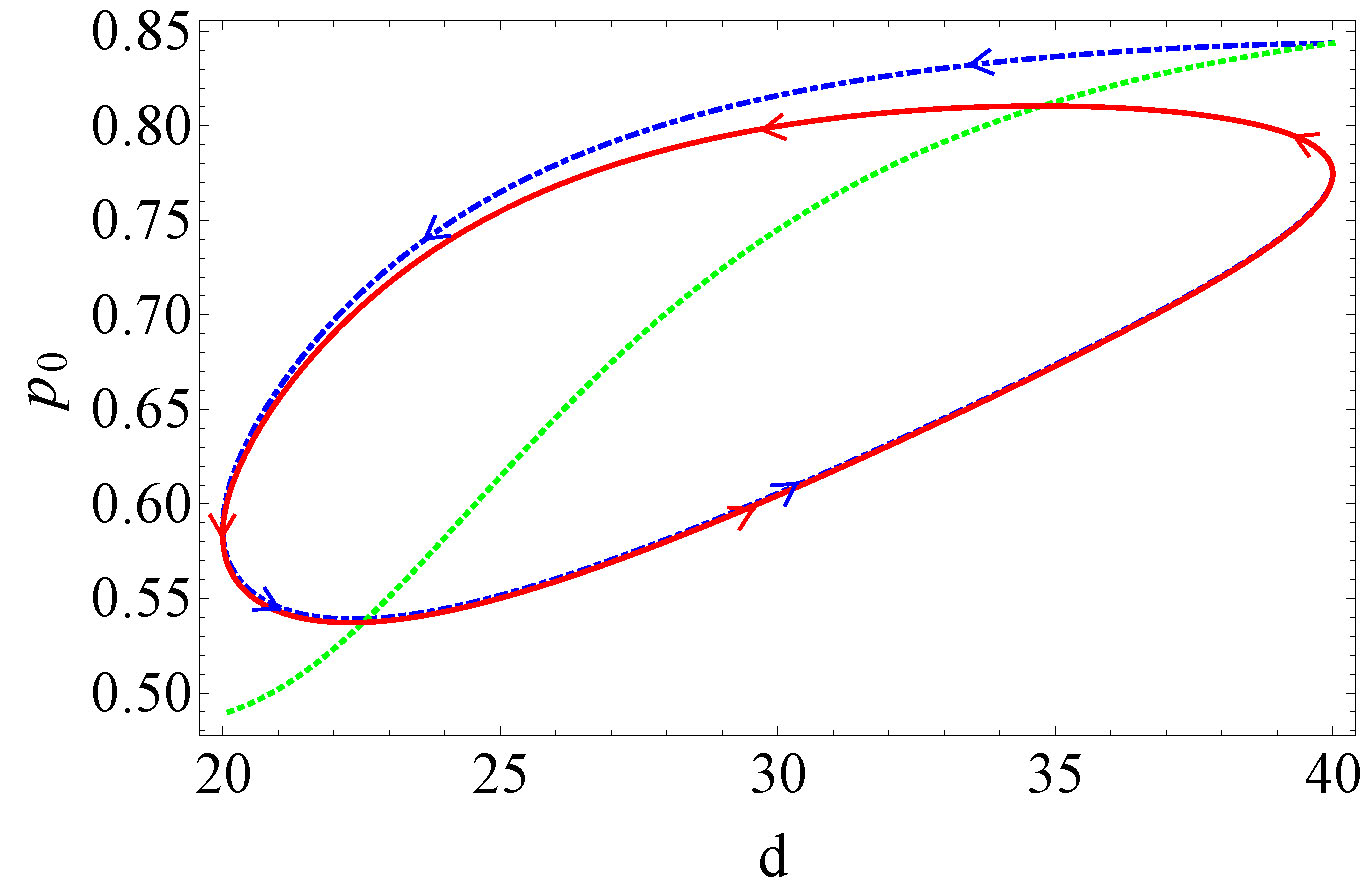}
\hspace{0.5cm}
\includegraphics[width=7cm]{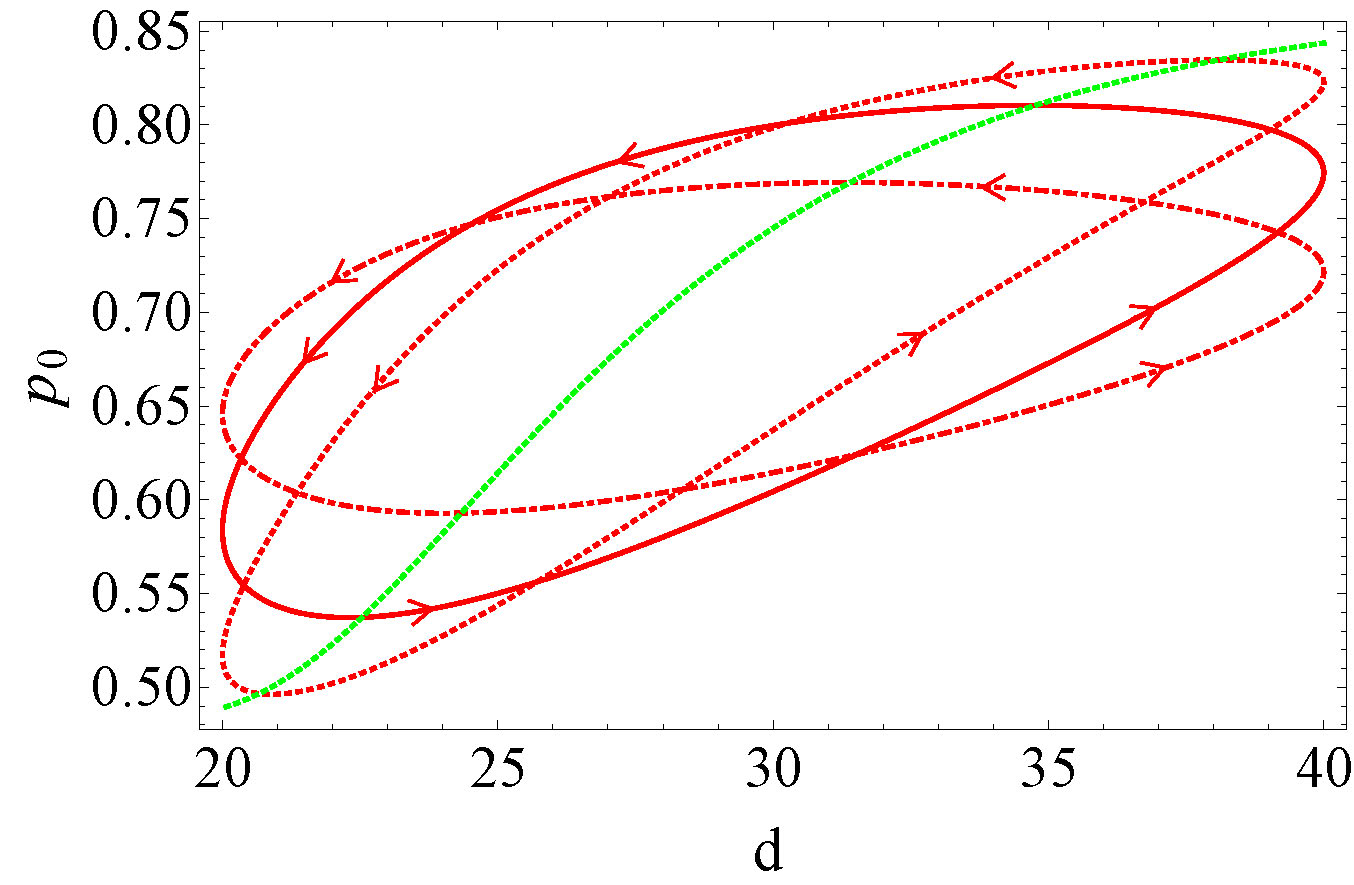}
\end{minipage}
\end{center}
\caption{Population $p_0$ of the instantaneous
ground state of $H_{M}(t)$ \textit{vs.} the molecular configuration $d$.
the oscillation parameters and the color code for the curves are the same as in
Fig.~(\ref{QSMEb}).
Here, the arrows along the curves indicate how the ground state population
evolves during the classical oscillations.}
\label{QSMEa}
\end{figure}

In Fig.~(\ref{QSMEa}) we plot the population of the instantaneous
ground state of the system Hamiltonian $H_{M}(t)$ during the
oscillation and observe how the periodic oscillations
keep the molecule far away from its thermal equilibrium (as compared
with the green curve).
In particular, when the spins get closer and the entanglement of the
ground state increases, the ground state occupation is noticeably higher
than the thermal one, and this actually constitutes the key of the mechanism.
Let us point out once more that thermalization tends to drive the
molecule close to its instantaneous equilibrium state, but
the continuous change of its shape prevents a full thermalization.
The intrinsic reset effect of environmental noise is exploited
to increase the ground state population when two spins reach the distant
configuration.
One can conclude that, even in presence of a very hot environment, there is
still a chance to see entanglement in an oscillating molecule after
an arbitrarily long time, see Fig.~(\ref{QSMEb}).

\subsection{Beat dephasing via thermalization}

\begin{figure}[bht]
\begin{center}
\begin{minipage}{15cm}
\includegraphics[width=7cm]{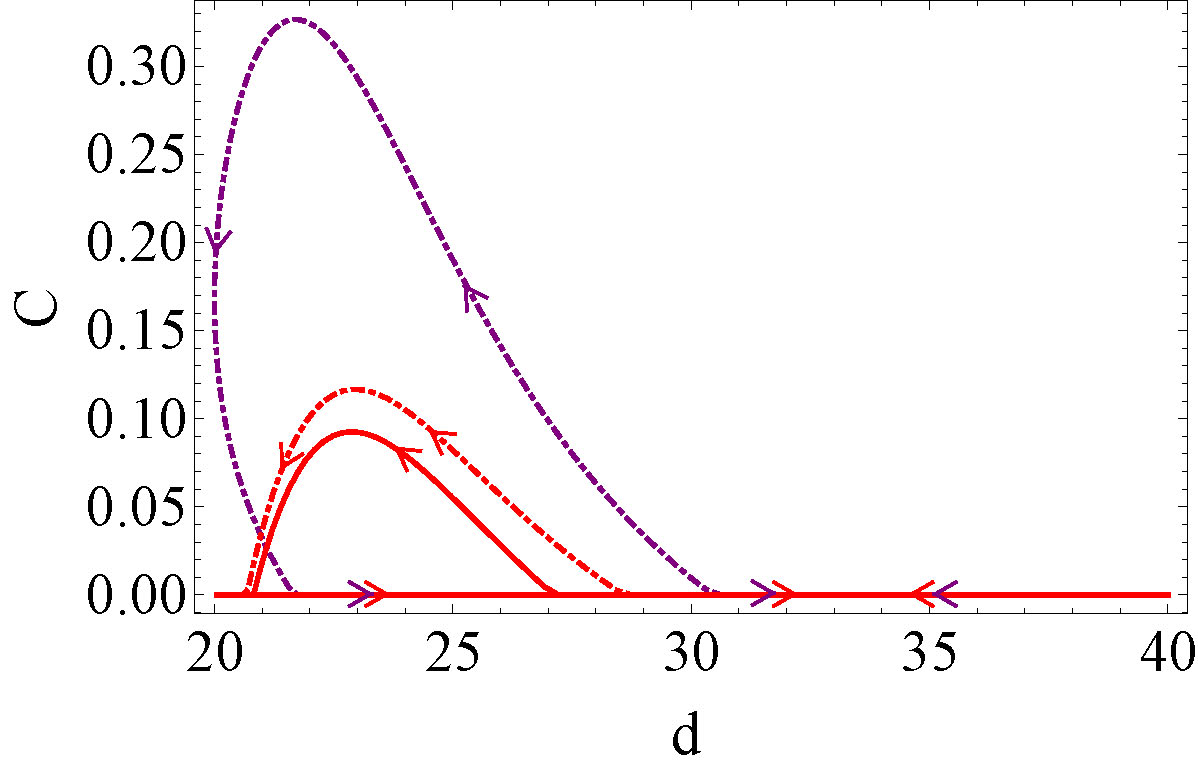}
\hspace{0.5cm}
\includegraphics[width=6.8cm]{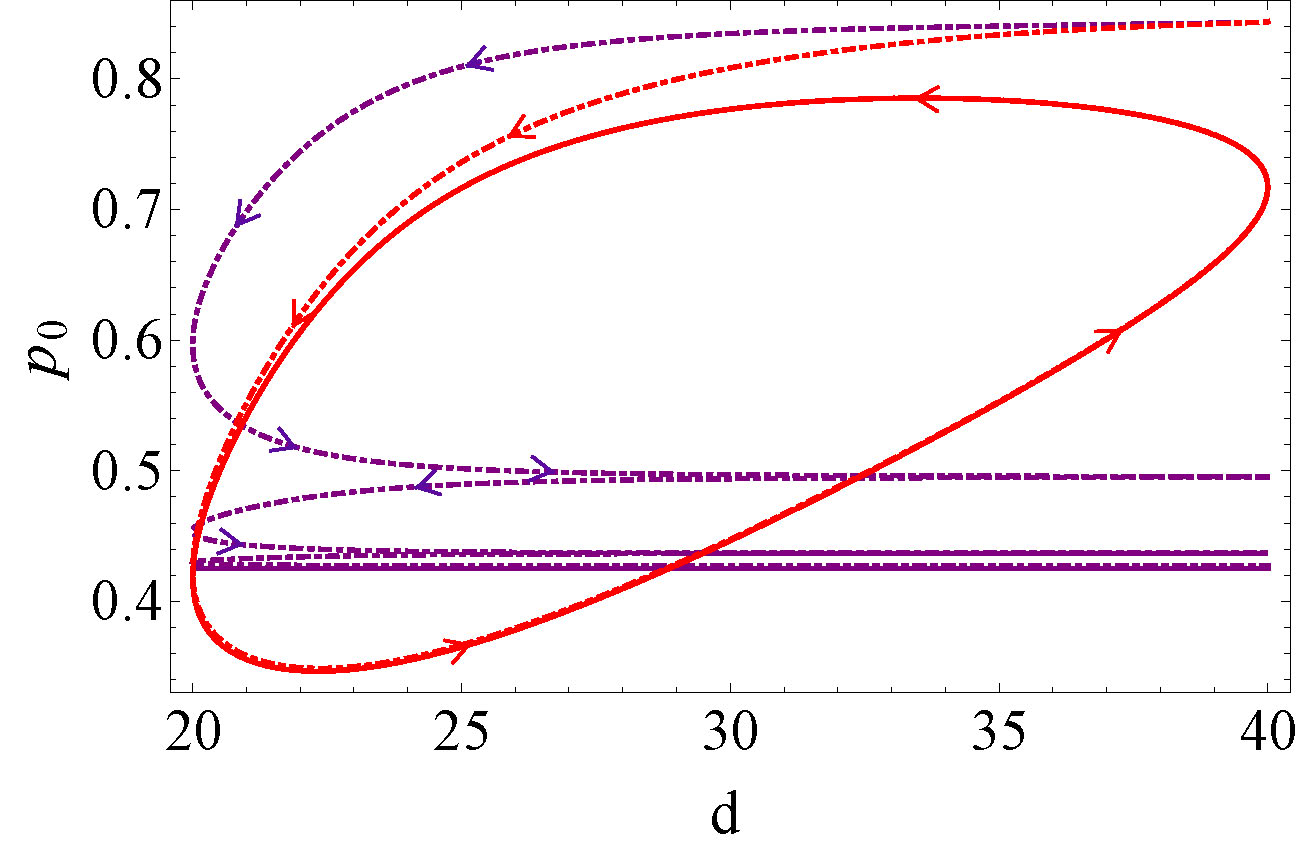}
\end{minipage}
\end{center}
\caption{Entanglement $C$ (left) and ground state population $p_{0}$ (right)
\textit{vs.} the molecular configuration $d$.
The dephasing rate is $\gamma_{p}=0.02$. The system-bath coupling strength
is $\kappa=0$ (purple curves) or $\protect\kappa=0.01$ (red curves).
The oscillation parameters are the same as Fig.~(\ref{QSMEb}) left.}
\label{BDT}
\end{figure}

The constructive role played by the environmental noise can be emphasized
by including an additional source of decoherence to further destroy entanglement.
Let us consider a ``phenomenological'' pure dephasing process acting
on the two spins via the term
\begin{equation}
\mathcal{D}_{p}\rho(t)=\gamma_{p}\sum\limits_{i=1}^{2}[\sigma_{z}^{(i)}%
\rho(t) \sigma_{z}^{(i)}-\rho(t)] ,
\end{equation}
with $\gamma_{p}$ denoting the dephasing rate.
As it is well known, pure dephasing destroys all the coherence
and always opposes the generation of entanglement.
Even for the oscillating molecule, the effect of pure dephasing alone is
to remove the off-diagonal density matrix elements of the two spin state,
and thereby destroy entanglement in the long run.
This is evident in Fig.~(\ref{BDT}), in which entanglement
can be found only in the first oscillation period and certainly not
in the asymptotic cycle (i.e. for $t\rightarrow \infty$).
What happens in the presence of a hot thermal bath?
One would expect that entanglement were destroyed faster due to this extra
decoherence source, and this is indeed true as long as only the first
oscillation period is taken into account, see Fig.~(\ref{BDT}).
However the constructive effect of thermalization becomes evident
in the successive cycles in which it partly counteracts the pure dephasing
through its built-in reset mechanism of re-pumping the ground state population
(as explained in the above subsection).

\subsection{Stochastic classical motion}
\label{SCM}

So far we have considered a deterministic and purely periodic classical motion.
In the context of biological systems, a more realistic approach
naturally includes a certain degree of stochasticity.
Different scenarios are conceivable depending on the biological
situations that one intends to model.

\subsubsection{Fluctuating configurations.}
The allosteric process depicted in Fig.~\ref{AllostericDevice} is
usually triggered by the interaction with a second chemical compound:
when the molecule is in the open configuration, a chemical docks on
the molecule and causes its allosteric contraction. The molecule reaches
its closed configuration and stays there until the chemical is released.
When the chemical undocks the molecule opens \cite{Alberts08,Hans09}.

\begin{figure}[hbt]
\begin{center}
\includegraphics[width=9.0cm]{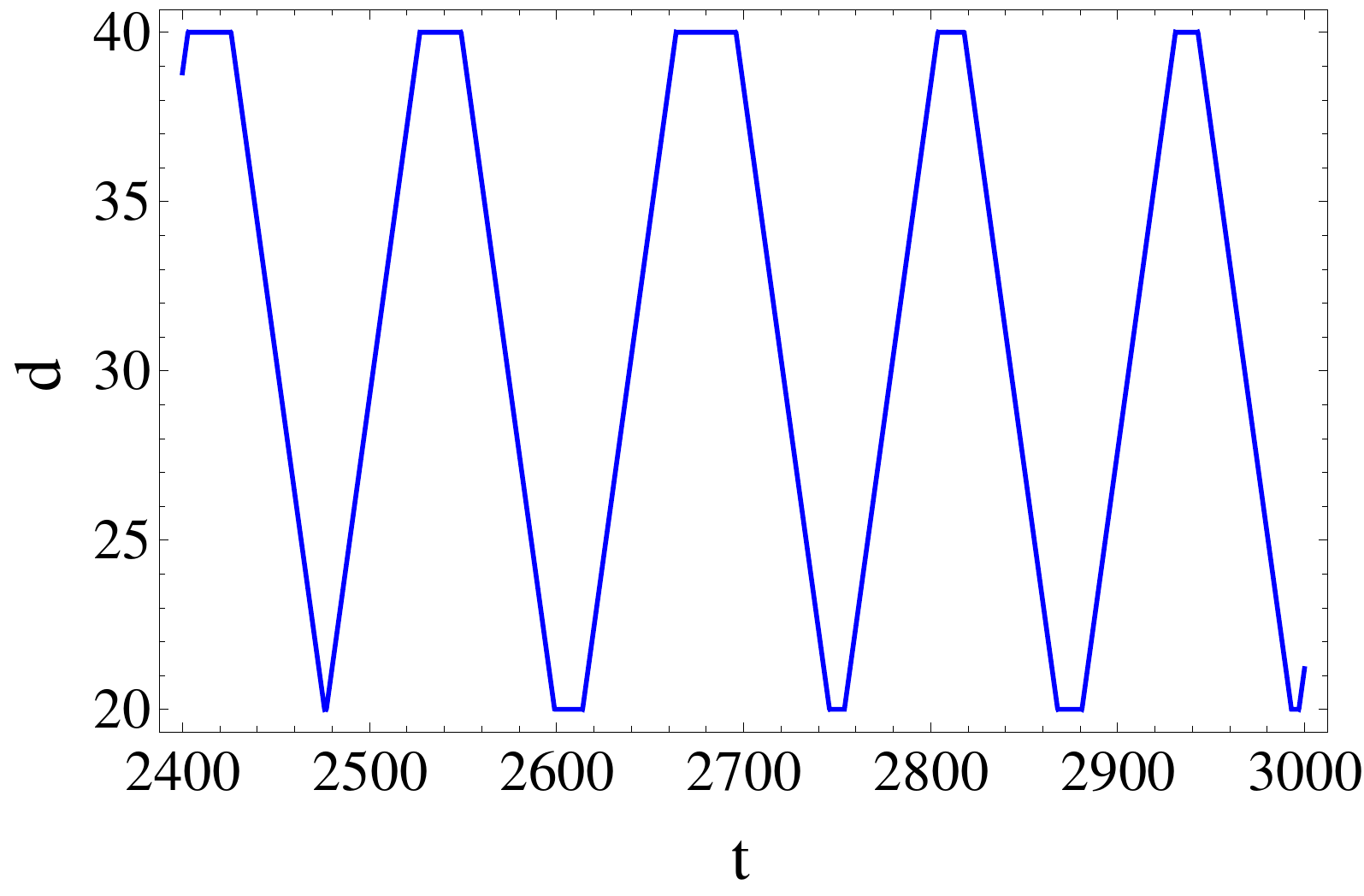}
\end{center}
\vspace{-.2cm}
\caption{Example of a trajectory for waiting times distributed according
to normal distributions (neglecting their tails for negative values).
Mean $\bar{d}$ and standard deviation $\sigma$ in the open (closed) configuration are
$\bar{d}=14.5$ , $\sigma=6$ ($\bar{d}=7.5$ , $\sigma=5$) respectively.}
\label{waiting}
\end{figure}

The waiting times in the open and closed configuration can be thought as two stochastic
variables which are, in principle, different.
Supposing normal (Gaussian), but independent distributions for both processes, one obtains trajectories
like the one depicted in Fig.~(\ref{waiting}), where the linear change in the spin distance during the
conformational process is an arbitrary, but nonessential choice.

Owing to our understanding obtained in the deterministic case,
the intuition correctly predicts the limiting cases of long and short waiting times:
when the chemical frequently docks or undocks from the molecule, the system is
continuously kept out of equilibrium and its dynamics closely resembles
the deterministic case.
On the other hand, for long waiting times (\emph{i.e.} long compared to the thermalization rate)
equilibrium is reached in the extreme configurations and during the contraction
of the molecule the \emph{single-shot} behavior is reproduced.
The creation of entanglement is here only a transient effect.

\subsubsection{Potential landscape and thermal vibrations.}\label{potential+vibrations}
\begin{figure}[hbt]
\begin{center}
\includegraphics[width=9.0cm]{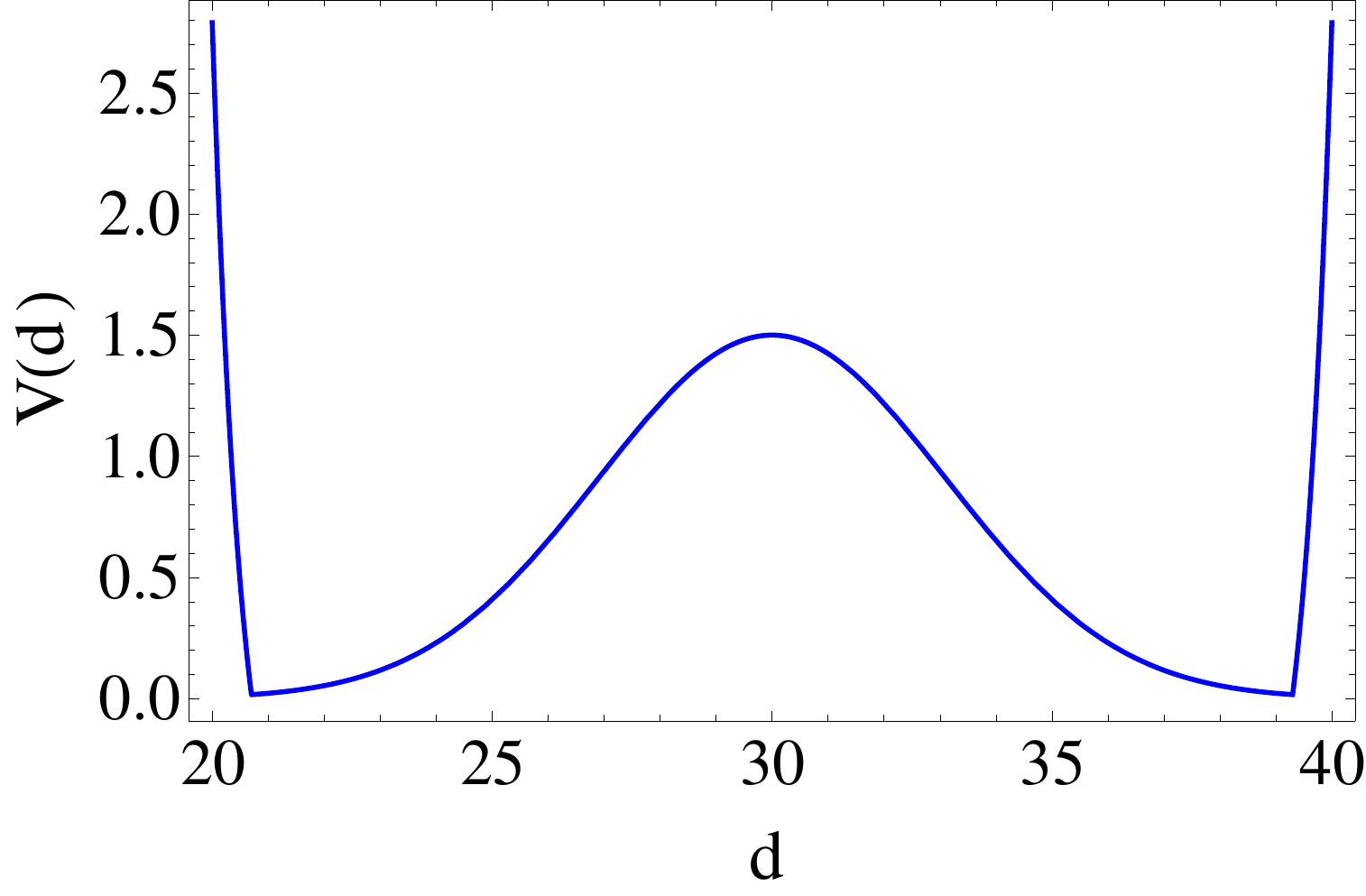}
\end{center}
\vspace{-.2cm}
\caption{Energy landscape $V(d)$ of molecular configurations with respect to the spin distance $d$.
For consistency, we also set the temperature of this environment to $\beta=1$ and choose the
barrier height accordingly.}
\label{landscape}
\end{figure}

A somewhat more interesting dynamics is obtained in the following scenario:
the molecule has two configurations which minimize the potential
energy associated to its classical shape.
In our model it is immediate to identify the distance between the spins with
an effective collective position coordinate for the macromolecule:
varying such coordinate we move from the low-energy closed configuration
to the low-energy open configuration via an energy landscape that contains
barriers.
For the sake of simplicity we model such energy surface as presenting only
one central barrier and we add two steep increasing gradients to forbid
configurations which are too extended or contracted, see Fig.~(\ref{landscape}).

\begin{figure}[thb]
\begin{center}
\begin{minipage}{15cm}
\hspace{-.4cm}
\includegraphics[width=7cm]{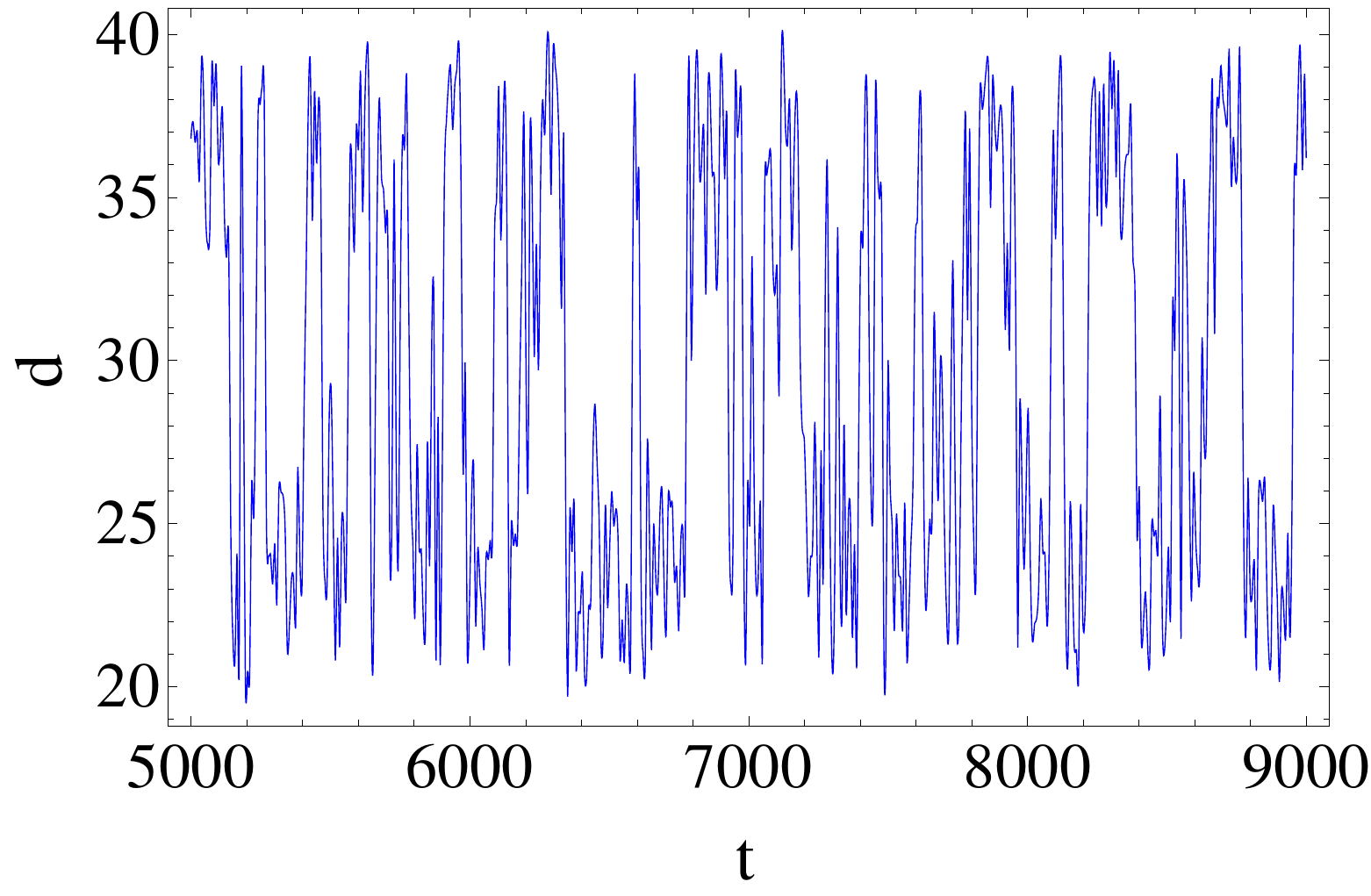}
\hspace{0.5cm}
\includegraphics[width=7cm]{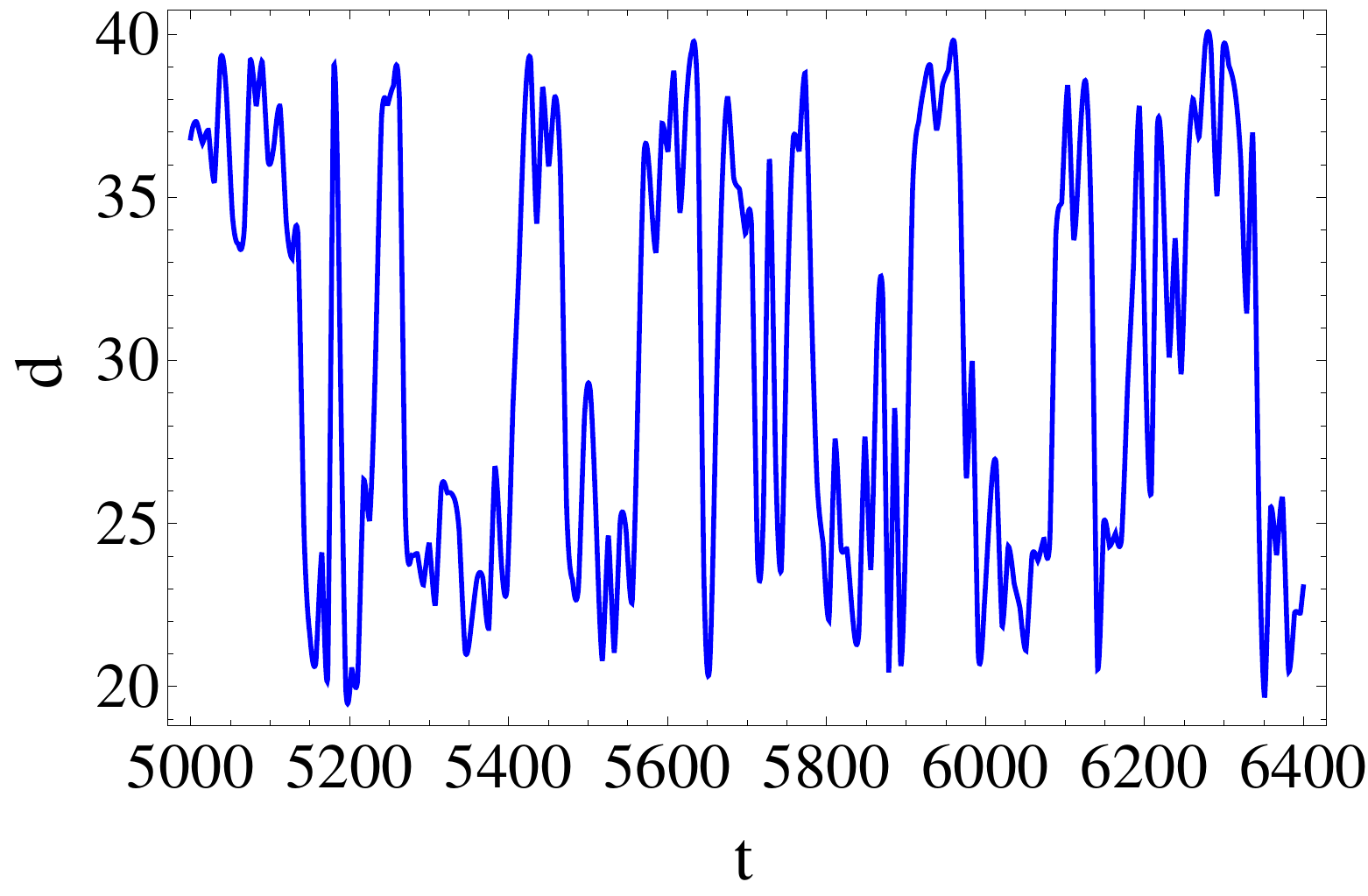}
\end{minipage}
\end{center}
\caption{Left: Example of a trajectory for the potential landscape
depicted in Fig.~(\ref{landscape}) once thermal fluctuations are taken into account.
Right: Magnification of a part of it.}
\label{potential}
\end{figure}

Once a configuration is chosen (as an arbitrary initial condition), the molecule starts
exploring the other possible configurations due to the thermal energy supplied by the
surrounding environment (notice that it is a totally different and unrelated environment
with respect to the bosonic heat bath).
It means that the molecule can randomly contract or stretch, as long as the energy of the
new configuration is within thermal fluctuations.
Using Metropolis algorithm and the potential landscape of Fig.~(\ref{landscape}),
we obtain trajectories as the one shown in Fig.~(\ref{potential}).

\begin{figure}[hbt]
\begin{center}
\includegraphics[width=9.0cm]{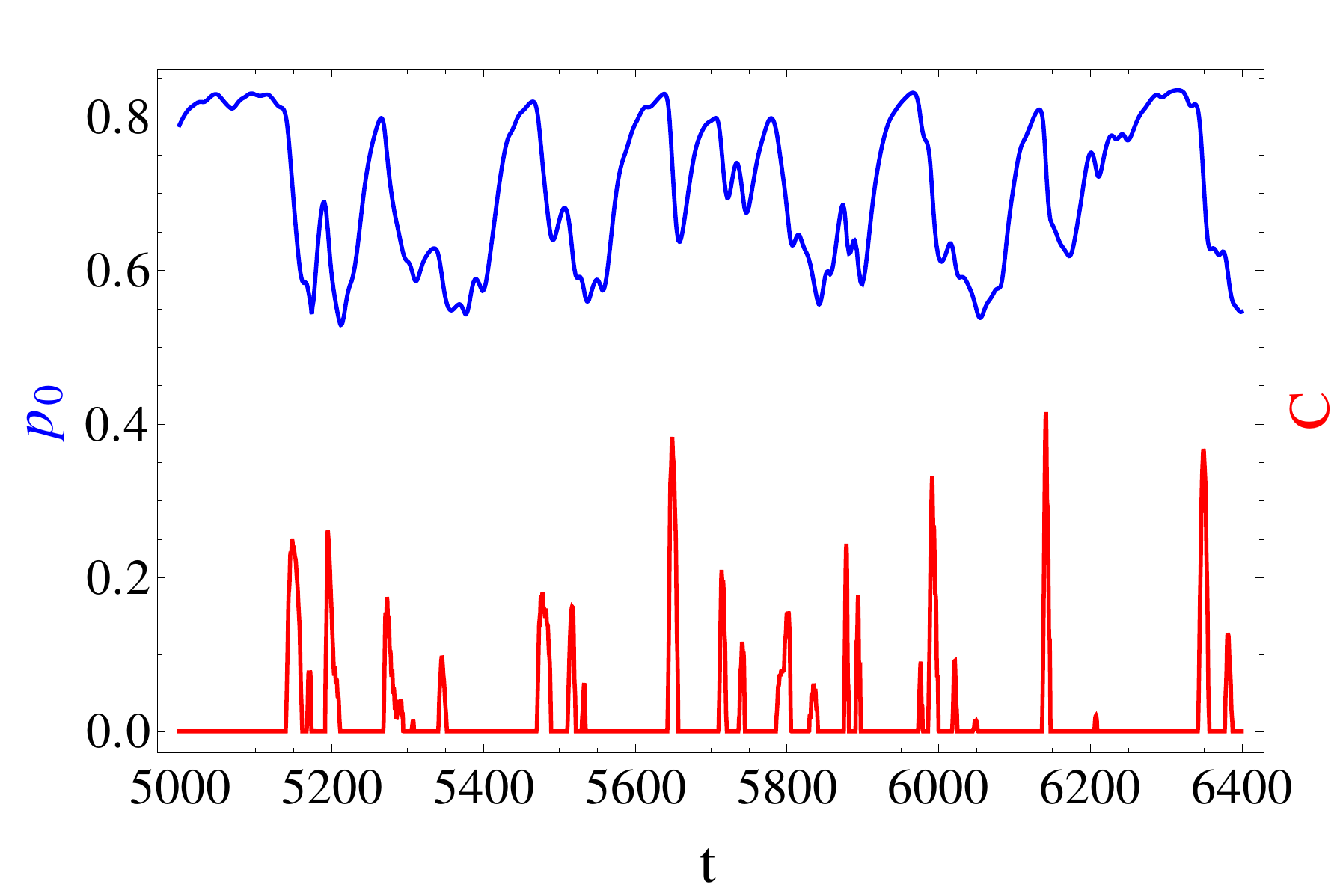}
\end{center}
\vspace{-.2cm}
\caption{Ground state (upper, blue) and concurrence (lower, red) in presence
of the thermally driven fluctuations of the molecule's configuration.}
\label{potential_results-2}
\end{figure}

In this situation, one cannot describe the molecular motion as
composed by distinct opening and closing processes.
In fact, the molecule stretches and contracts its configuration
only partially and in a stochastic way.
Nevertheless, the mechanisms presented in subsection~\ref{Persistent recurrence of fresh entanglement}
are still at work in this scenario.
Due to the increasing energy gap, the ground state population raises when
the distance between the spins increases, with an ``inertial'' reaction time
determined by the thermalization rate.
Whenever the system undergoes a relatively fast contraction \emph{after}
having spent enough time in an extended configuration, entanglement
is generated, as clearly appears in Fig.~(\ref{potential_results-2}).

% From JM's paragraph on average entanglement:
% It can be seen that the persistent generation of dynamic entanglement
% can appear in a hot environment not only from period oscillation,
% but also from more realistic stochastic classical motion.

\subsection{Non-Markovian case}

The microscopic derivation of the Lindblad type master equation previously
introduced in this section relies on several approximations
(for a more detailed derivation see \cite{Cai10}).
These approximations refer to assumptions regarding the time scale
of the thermal bath's dynamics and the strength of its interaction with the system.
In particular, such interaction is supposed to be weak and the bath's relaxation
time very short compared to the time scale of the system dynamics.
For many biological processes, these conditions may not be satisfied.
Nevertheless they can be considered as useful working hypothesis
to obtain analytical expressions (suitable for numerical integration),
and to illustrate our main idea.

To demonstrate that such simplifying assumptions are indeed not essential for the
validity of our arguments, we now extend our analysis to situations
with a not-so-weak system-bath interaction and a non-Markovian reduced dynamics.
We rephrase our problem within the path integral formulation of quantum mechanics:
In this formulation, the actual evolution of a quantum state is determined
by the coherent sum of all the possible evolutions weighted via a phase
proportional to the corresponding (classical) action.
The presence of an environment in contact with the system leads to an additional
weighting factor, the so-called influence functional \cite{FeynmanBook,Feynman63}.
The influence functional depends on the system's possible evolution and on the properties
of the bath, and in general is non-local in time.
This formalism is particularly convenient in case of environments composed of
non-interacting harmonic oscillators, where one can obtain an analytical
expression for the influence functional \cite{Feynman63}.

The part of the summation involving the system's degrees of freedom
has to be performed numerically, and a particularly favourable
prescription was introduced by Makri \& Makarov
under the acronym of QUAPI (QUasi Adiabatic propagator Path Integral)
\cite{Makri,Thorwart11}.
Their method allows one to include a finite time memory via a truncated influence
functional which takes into account only a few of the last propagation steps.
The memory time is determined by how many time steps are included and by how
long each of them lasts: $T_{mem}=\Delta k \, \Delta t$.
In principle, the longer $T_{mem}$ the further one can relax the Markovian assumption.
In practice, however, the computational effort scales exponentially with $\Delta k$
and the error in the single-step propagation quadratically in $\Delta t$.
To overcome these limitations, each simulation has to be performed with different
parameters and one has to perform two extrapolation procedures:
the first keeps constant $T_{mem}$ while $\Delta t \rightarrow 0$ to obtain
precise estimate with finite memory,
the second increases $T_{mem} \rightarrow \infty$ to extrapolate the
infinite memory time limit.

\begin{figure}[htb]
\begin{center}
\begin{minipage}{14cm}
\includegraphics[width=6.3cm]{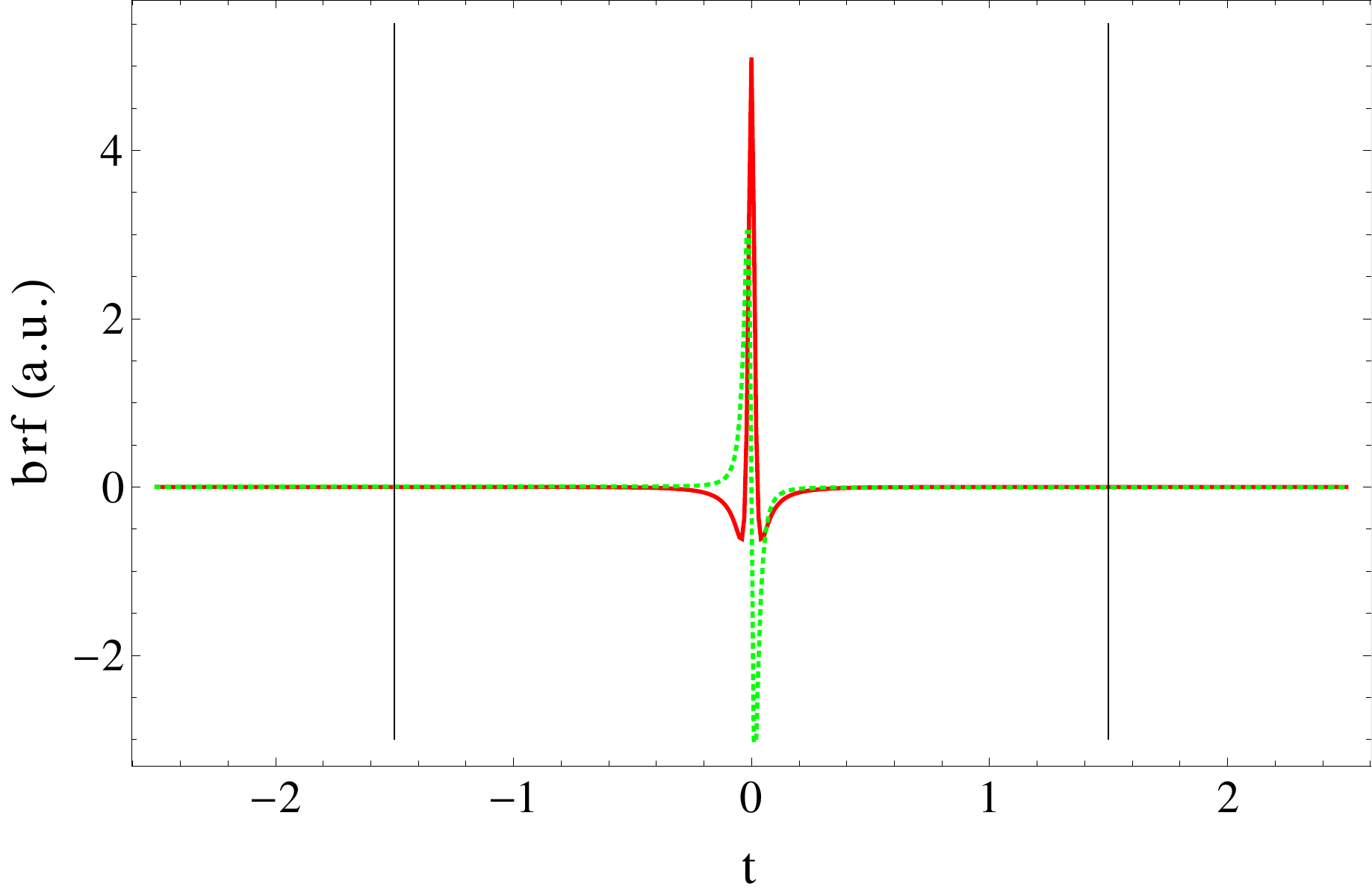}
\hspace{0.5cm}
\includegraphics[width=6.7cm]{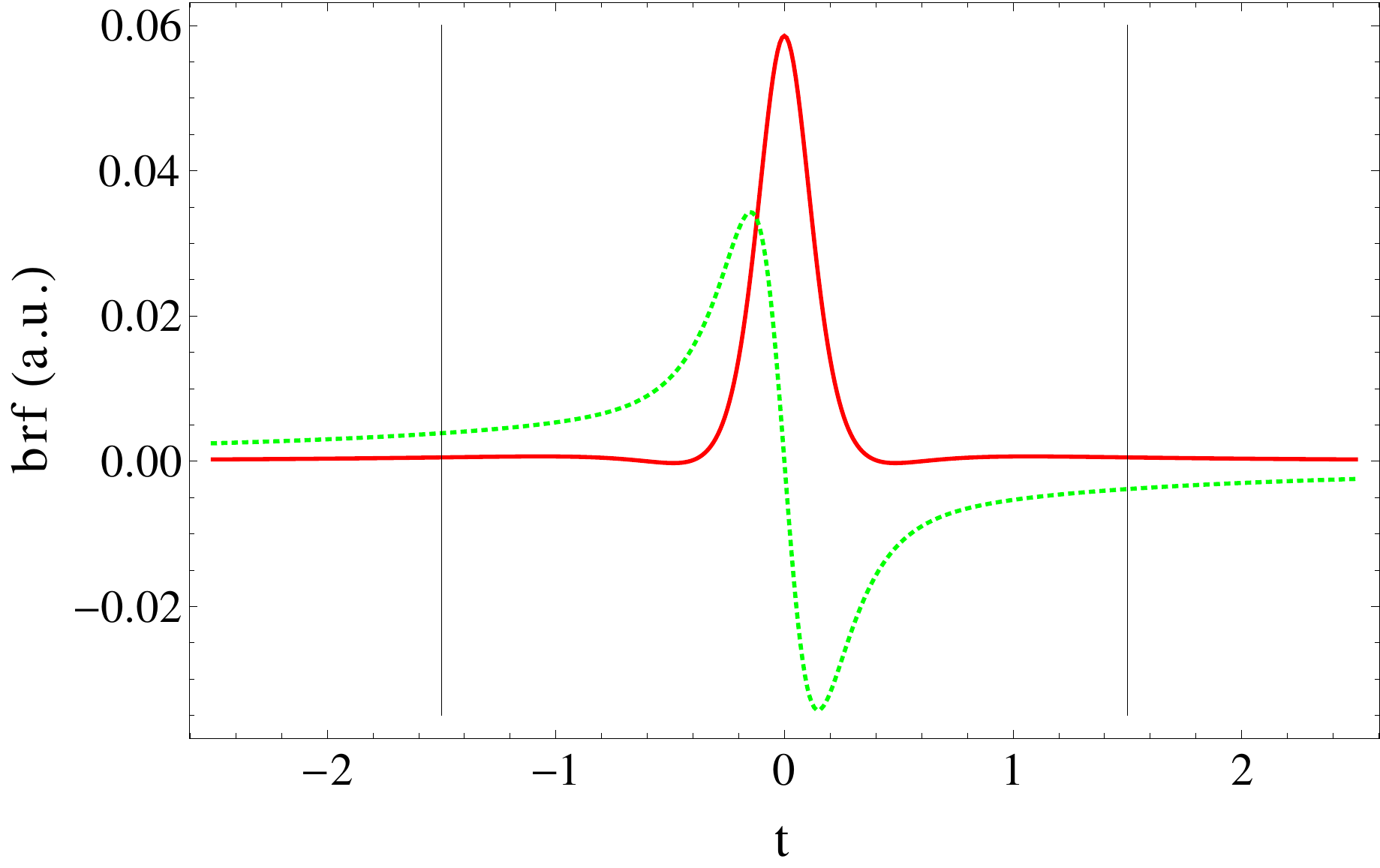}
\end{minipage}
\end{center}
\caption{Bath response function for Markovian ($\omega_c=40$, top)
and non-Markovian ($\omega_c=4$, bottom) regimes.
The solid, red curve is the real part of the BRF, the dashed, green curve is
its imaginary part, while the vertical, black lines mark the memory time
$T_{mem}=1.5$ considered in Fig.~(\ref{QUAPI}).}
\label{brf}
\end{figure}

\begin{figure}[htb]
\begin{center}
\begin{minipage}{12cm}
\includegraphics[width=5.5cm]{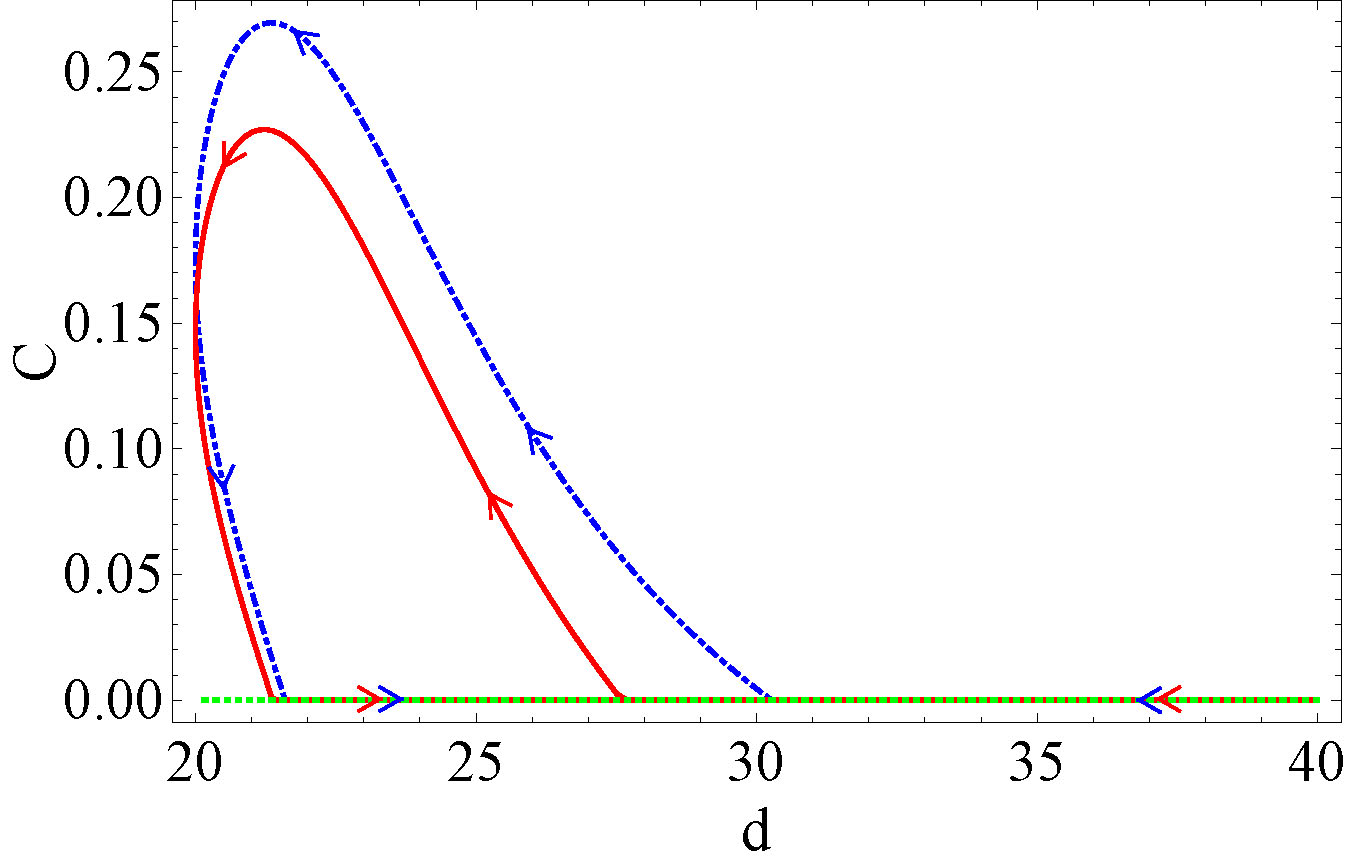}
\hspace{0.5cm}
\includegraphics[width=5.5cm]{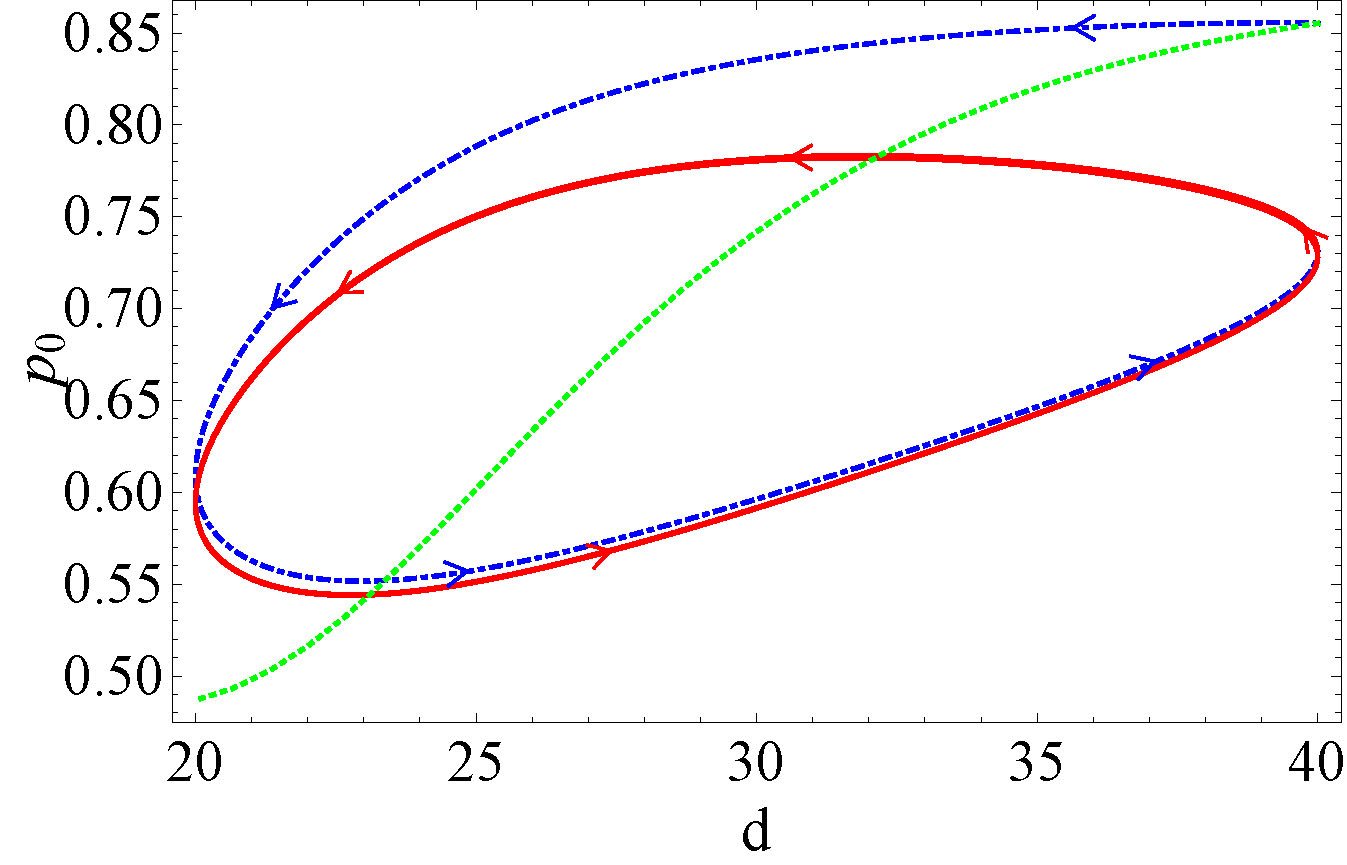}
\end{minipage}
\begin{minipage}{12cm}
\includegraphics[width=5.5cm]{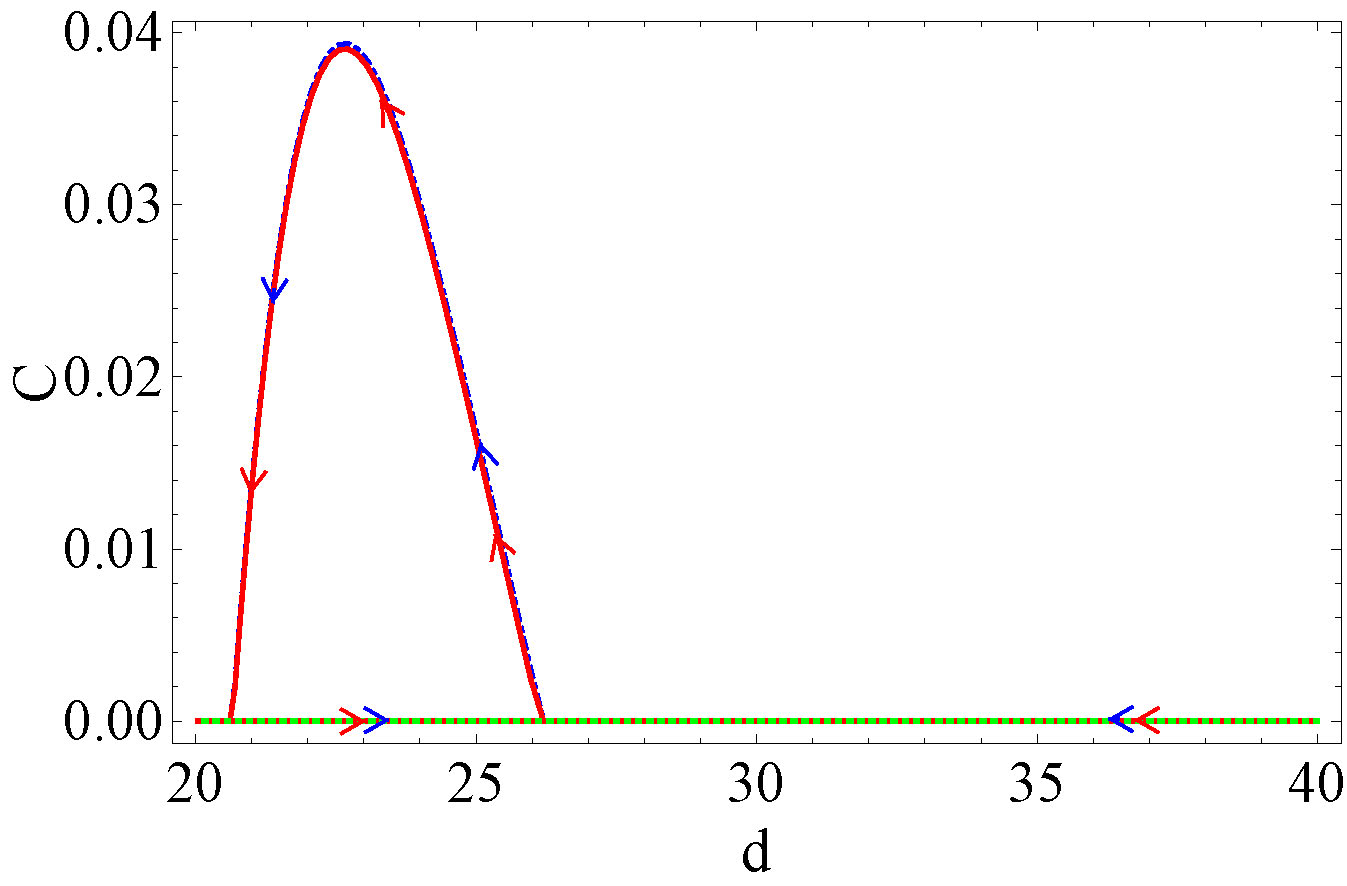}
\hspace{0.5cm}
\includegraphics[width=5.5cm]{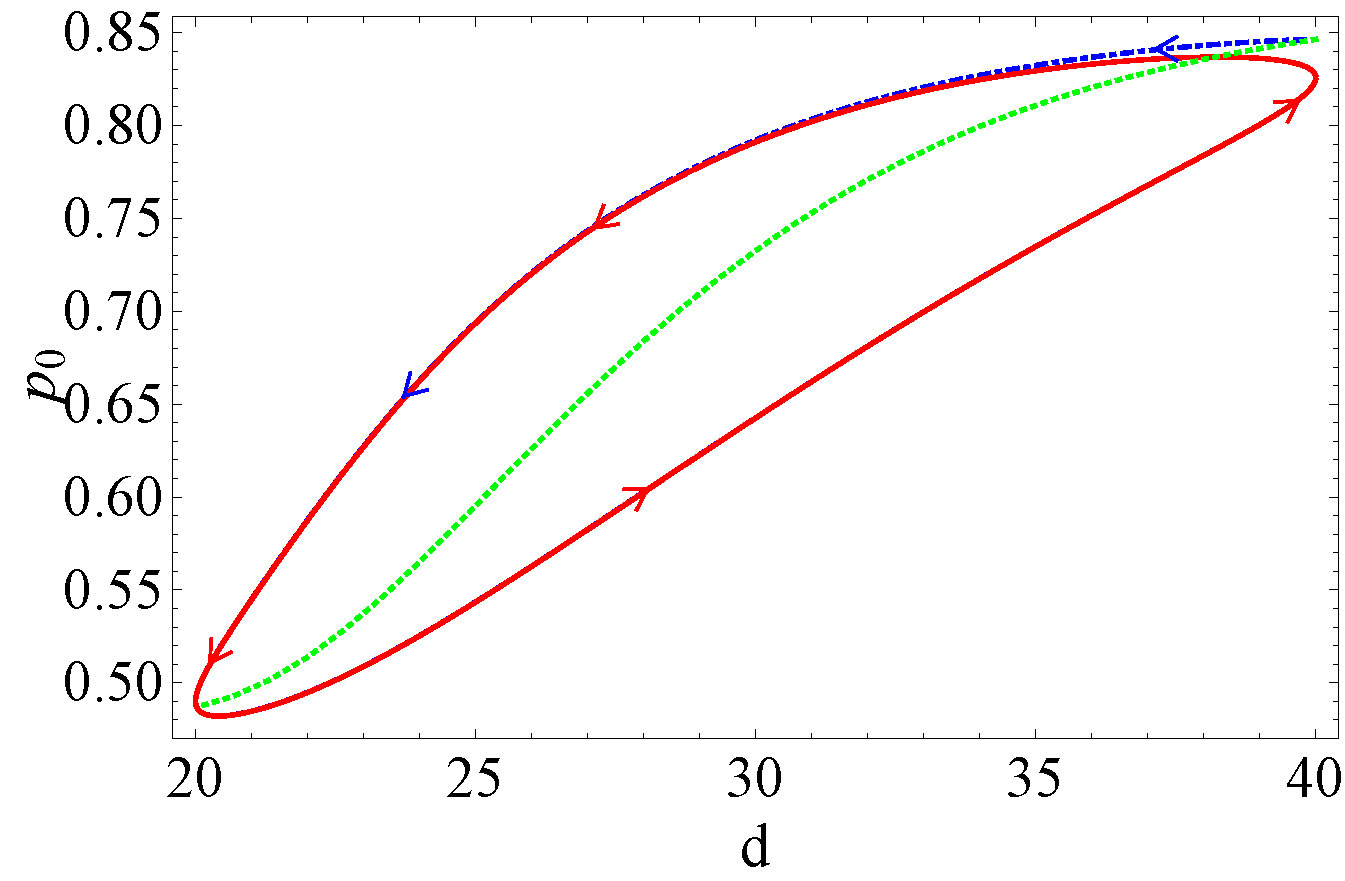}
\end{minipage}
\end{center}
\caption{Entanglement $C$ (left) and ground state population $p_0$ (right)
\textit{vs.} molecular configuration $d$ for a non-Markovian environment.
The dot-dashed, blue curve indicates the first period;
the solid, red curve marks the asymptotic cycle;
while the green, dotted curve refers to the QUAPI equilibrium state for each specific configuration.
Top: $\omega_c=4 , \kappa=0.01 , T_{mem}=1.5 , \tau=100 , \Delta t=5$;
Bottom: $\omega_c=4 , \kappa=0.05 , T_{mem}=1.5 , \tau=75 , \Delta t=5$;
while the other parameters are as in Fig.~(\ref{QSMEb}) left.}
\label{QUAPI}
\end{figure}

These technical procedures are extensively presented in \cite{Thorwart06}
and are beyond the aim of this work:
Here we do not want to precisely quantify the amount of entanglement generated
under a specific system-environment interaction, but our goal is to demonstrate
that the mechanism we introduced is indeed robust and is not confined
to the Born-Markov regime.
However, we have selectively performed the extrapolation procedures for several
points along the curves: The agreement is always good (on the \% level)
and consistent with the values of entanglement found before the extrapolations.
The extent of the memory effects is determined by the so called bath response function BRF:
To properly take into account the non-Markovianity of the environment, the memory time
$T_{mem}$ has to be larger than the bath correlation time given by the width of the BRF,
see Fig.~(\ref{brf}).
Fig.~(\ref{QUAPI}) shows how the generation of entanglement persists even
for baths with finite-time correlations (determined by a small value for the cut-off
frequency in the spectral density $J(\omega)$, giving rise to finite memory effects)
and moderately strong system-environment couplings.

\section{Spin gas model}
\label{SpinGas}

The actual environment for biological systems is expected to be very
specific and complex, and has not yet been fully characterized.
In the above section we investigated the case of thermal baths
composed of a large number of harmonic oscillators, but this is most
probably still a simplistic (even if reasonable and common) choice.
In this perspective, it is worth to see whether the main features
that we observe for bosonic heat baths are preserved also for a
different kind of environment, namely the spin gas model
\cite{Cals05,Lor06,Lor07}, which is particularly suggestive
in a biological context.
Here the environment is composed by particles which follow random
classical trajectories and collide with our two spins.
During the collisions they stochastically interact with the spins
giving rise to an effective decoherence process.
Under certain assumptions \cite{Lor07}, at a coarse time level
that includes both the coherent Hamiltonian dynamics and the random
collisions, it is possible to derive a phenomenological description
in terms of a Lindblad-type quantum master equation characterized
by the following dissipator:
\begin{eqnarray}
\mathcal{D}\rho(t) =\gamma \sum_{\alpha=1}^{2} [\mathcal{L}
_{g}^{(\alpha)}\rho(t)+ \mathcal{L}_{d}^{(\alpha)}\rho(t) ] ,
\label{ME1}
\end{eqnarray}
with
\begin{eqnarray}
\mathcal{L}_{g}^{(\alpha)}\rho &=&s[2\sigma _{+}^{(\alpha)}\rho \sigma
_{-}^{(\alpha)}-\rho \sigma _{-}^{(\alpha)}\sigma _{+}^{(\alpha)}-\sigma
_{-}^{(\alpha)}\sigma _{+}^{(\alpha)}\rho ] \\
\mathcal{L}_{d}^{(\alpha)}\rho &=&(1-s)[2\sigma _{-}^{(\alpha)}\rho
\sigma _{+}^{(\alpha)}-\rho \sigma _{+}^{(\alpha)}\sigma
_{-}^{(\alpha)}-\sigma _{+}^{(\alpha)}\sigma _{-}^{(\alpha)}\rho ] ,
\end{eqnarray}
where $\sigma_{\pm}^{(\alpha)}=(\sigma_{x}^{(\alpha)}\pm i
\sigma_{y}^{(\alpha)})/2$ are the ladder operators for a two-level
system. The de-coherent channels $\mathcal{L}_{g}^{(i)}(\rho)$,
$\mathcal{L}_{d}^{(i)}(\rho)$ represent the energy gain and loss
during the process: $\gamma$ is the rate with which $\langle (1+\sigma_z)/2 \rangle$ decays,
and $s$ is its equilibrium value \cite{Hans93}.
Without loss of generality, we assume that $\gamma >0 $ and $0 \leq s\leq 1/2$.
We also note that correlations in the spin gas can be included in the analysis
\cite{Cals07}.

\subsection{Static entanglement in the steady state}

\begin{figure}[htb]
\begin{center}
\includegraphics[width=8cm]{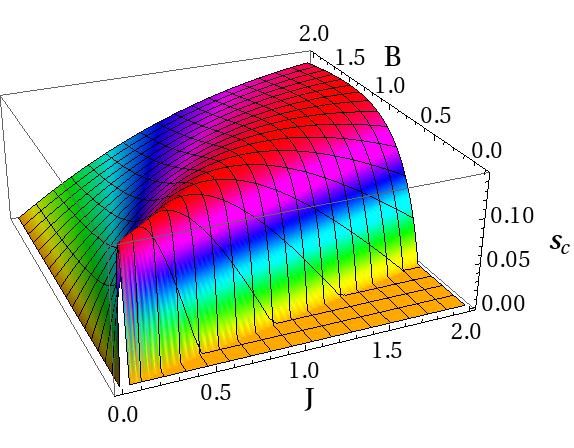}
\end{center}
\caption{The critical value of $s$ (above which no static
entanglement exists in the steady state) for different values of $B$ and $J$
with $\gamma=0.025$.} \label{CS}
\end{figure}

Similar to the case of the bosonic heat baths, we first consider the steady
state $\varrho \equiv \rho(t\rightarrow \infty)$ for each molecular
configuration, \emph{i.e.} at fixed spin-spin interaction strength $J$ and
background field $B$, to verify that it is a separable state.
To this end, we impose
\begin{equation}
-i[H_{M}(t),\rho]+\mathcal{D}\rho=0 .
\end{equation}
It is straightforward to check that the steady state for each configuration $d$
is unique and that its non zero density matrix elements are given by
\begin{equation}
\left\{
\begin{array}{l}
\varrho_{00}=[J^{2}+4 b^2 s^{2}]/4(J^{2}+b^2) \\
\varrho_{11}=[J^{2}+4 b^2s(1-s)]/4(J^{2}+b^2) \\
\varrho_{22}=[J^{2}+4 b^2s(1-s)]/4(J^{2}+b^2) \\
\varrho_{33}=[J^{2}+4 b^2(1-s)^{2}]/4(J^{2}+b^2) \\
\varrho_{03}=-J(2B-i\gamma)(1-2s)/2(J^{2}+b^2) \\
\varrho_{30}=-J(2B+i\gamma)(1-2s)/2(J^{2}+b^2)
\end{array}
\right. ,
\label{steadystate}
\end{equation}
where $b=(4B^2+\gamma^{2})^{1/2}$. The static entanglement of the steady
state quantified by concurrence is calculated as $C(\varrho)=2\max\{0,|%
\varrho_{03}|-(\varrho_{11} \varrho_{22})^{1/2}\}$, resulting in
\begin{equation}
C(\varrho)=2\max\{0,\frac{2 J b (1-2s)-[J^{2}+4 b^{2}(1-s)s]}{4( J^{2}+b^{2})}\} .
\end{equation}
Thus, no static entanglement is possible if the parameter $s$
exceeds the critical value of
\begin{equation}
s_c \equiv \max\{0,\frac{1}{2}[1+J/b-(1+2 J^{2}/b^{2})^{1/2}]\}
\label{SC}
\end{equation}
plotted in Fig.~(\ref{CS}).

\subsection{Persistent dynamic entanglement}

The physical picture underlying the spin gas model is very different from
the collection of harmonic oscillators constituting the bosonic thermal bath
and these two environments destroy entanglement according to different mechanisms.
Nevertheless, they share the essential features which allow the cyclic
generation of fresh entanglement via the constructive role played
by the environment noise itself.
Even if in any possible static configuration the steady states for both models
and high enough temperature (bosonic heat baths) or level of noise (the spin
gas environment with a large value of $s$) does not exhibit entanglement,
it persistently appears when classical oscillations are taken into account.

\begin{figure}[htb]
\begin{center}
\begin{minipage}{15cm}
\includegraphics[width=7cm]{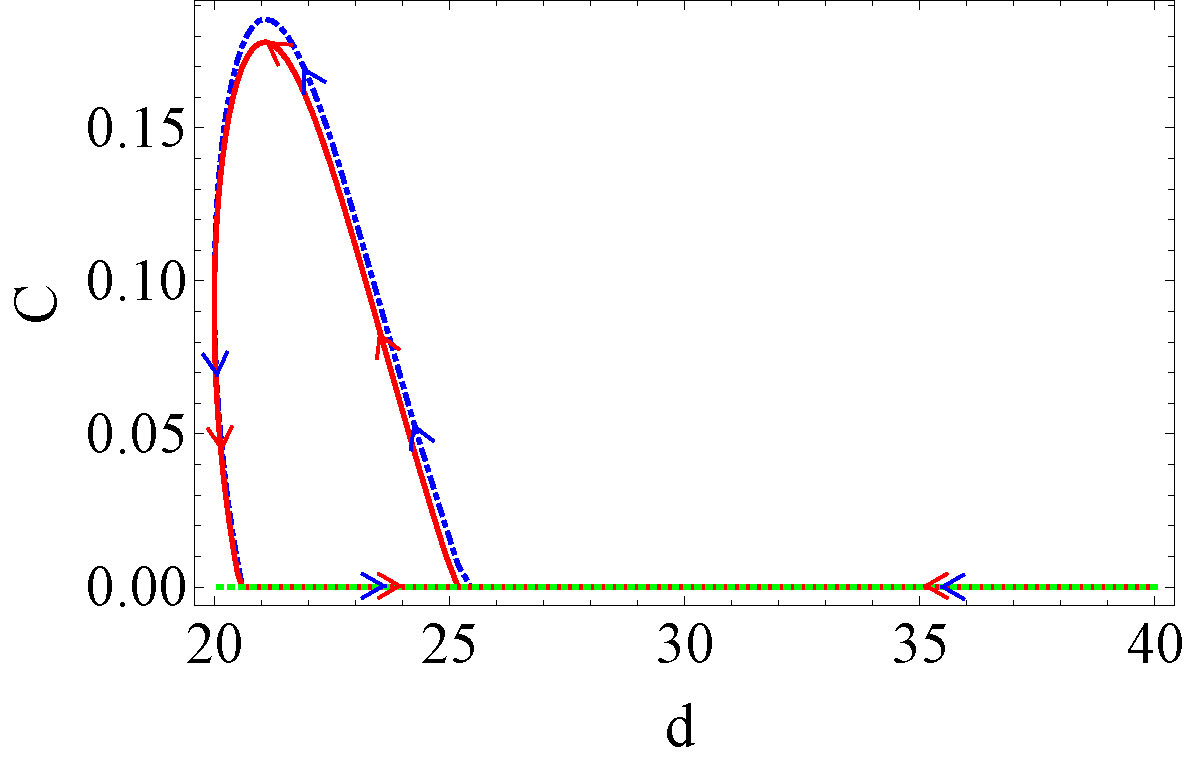}
\hspace{0.5cm}
\includegraphics[width=7cm]{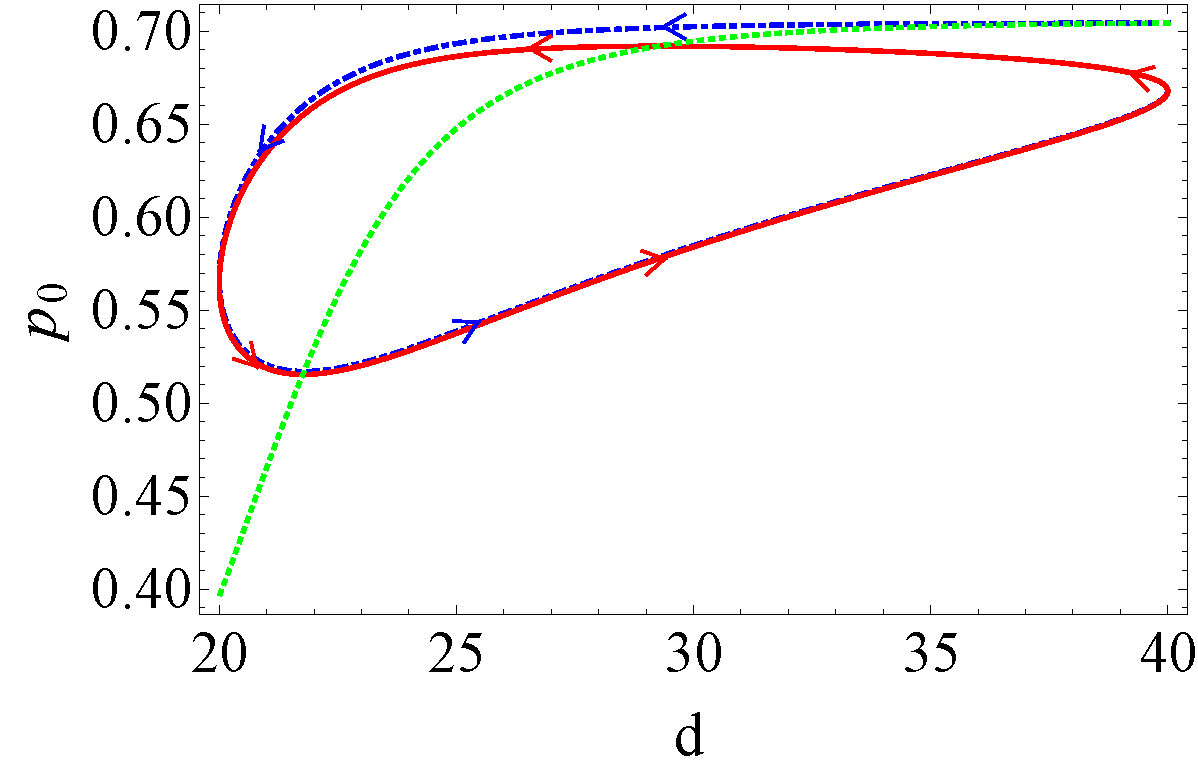}
\end{minipage}
\end{center}
\caption{Entanglement $C$ (left) and ground state population $p_{0}$
(right) \textit{vs.} the molecular configuration $d$ for the spin gas environment.
The dot-dashed, blue curve indicates the first period;
the solid, red curve marks the asymptotic cycle;
while the green, dotted curve refers to the steady state for a specific configuration.
The decoherence rates are $s=0.16$ and $\gamma=0.025$;
the oscillation parameters are the same as in Fig.~(\ref{QSMEb}) left.
}
\label{SPING}
\end{figure}

This can be explained since the spin gas environment also contains a built-in reset
mechanism via spin flip:
For the molecular configuration in which the two spins are spatially separated,
the environment effectively lowers the system's entropy and increases the ground state population.
Thus, under these conditions and in accordance with what has been observed for thermal baths,
dynamic entanglement is generated in the long time limit, Fig.~(\ref{SPING}).

\begin{figure}[htb]
\begin{center}
\begin{minipage}{15cm}
\includegraphics[width=7cm]{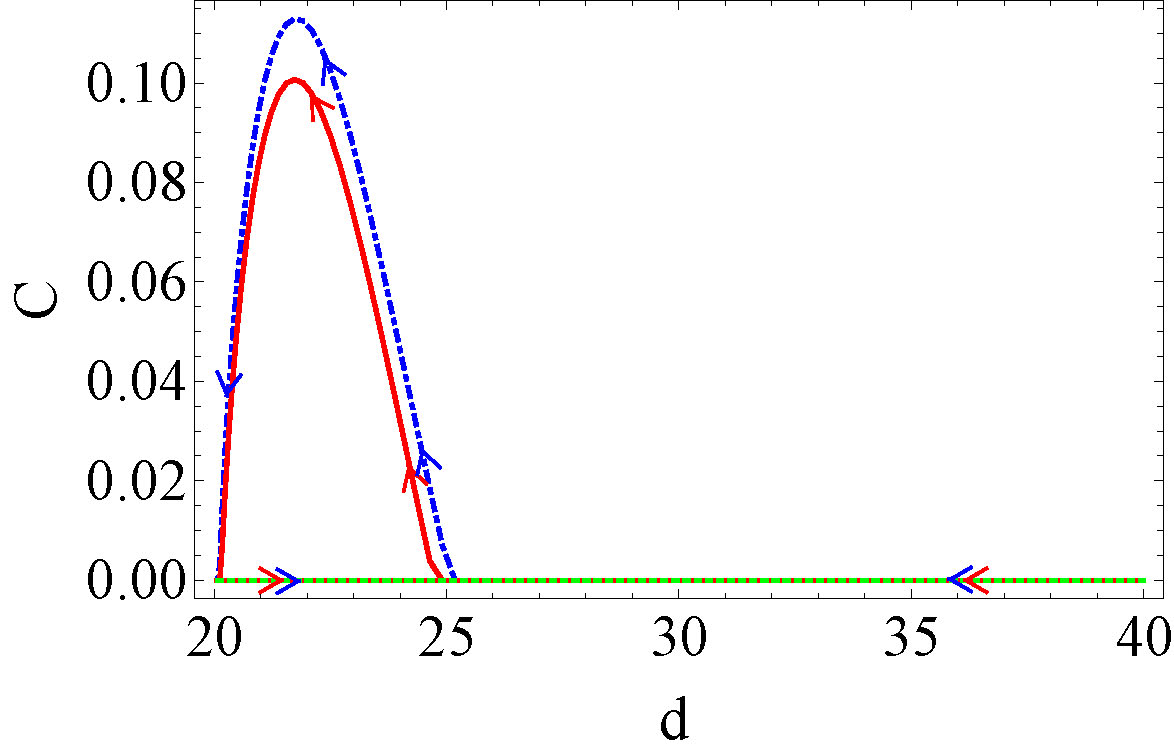}
\hspace{0.5cm}
\includegraphics[width=7cm]{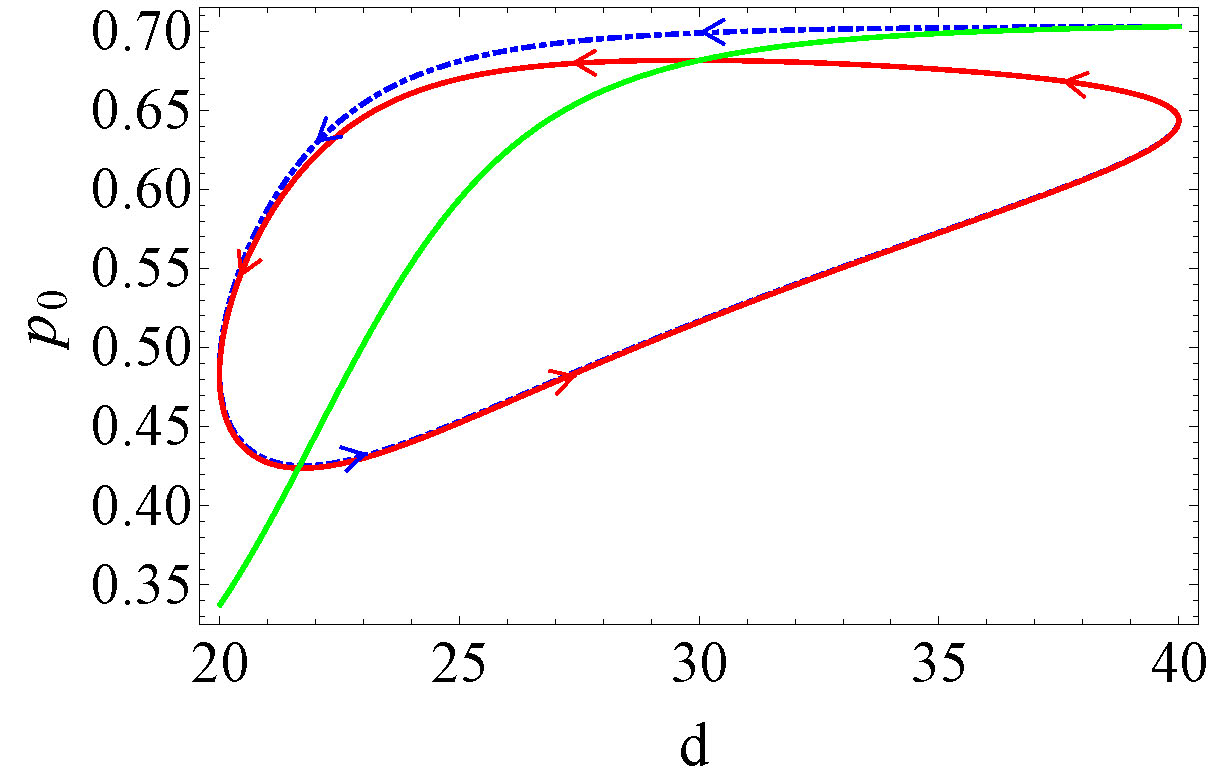}
\end{minipage}
\end{center}
\caption{Entanglement $C$ (left) and ground state population $p_{0}$
(right) \textit{vs.} the molecular configuration $d$ for the spin gas
environment with additional pure dephasing.
The dot-dashed, blue curve indicates the first period;
the solid, red curve marks the asymptotic cycle;
while the green, dotted curve refers to the steady state for a specific configuration.
The decoherence rates are $s=0.16$, $\gamma=0.025$, and $\gamma_{p}=0.01$;
the oscillation parameters are the same as in Fig.~(\ref{QSMEb}) left.
}
\label{SPINGDP}
\end{figure}

As in the case of the bosonic heat baths, here the constructive role of the
environment noise can also be further illustrated by adding pure dephasing.
One can see from Fig.~(\ref{SPINGDP}) that the noise in the spin gas model can
counteract the effect of pure dephasing and sustain the persistent dynamic entanglement.
The green curve is here obtained from an expression analogous to Eq.(\ref{steadystate}),
and generalized to include the additional dephasing.

\section{Concluding remarks}
\label{Conclusion}

In this work, we have demonstrated how entanglement can persist as a recurrent
property even in systems exposed to a hot and noisy environment.
To achieve this task we have exploited the presence of an internal driving force
which keeps the system out of the corresponding equilibrium state.
In biologically inspired context, such driving force can be identified with
conformational changes, including the stretching of macromolecules or
their folding/unfolding processes.
Here, we describe these conformational changes as purely classical, but
capable of influencing the dynamics of localized quantum degrees of freedom
and to prevent complete thermalization.
The non-equilibrium interplay between classical motion and open quantum
system dynamics gives rise to an effective reset mechanism
which promotes the generation of fresh entanglement even for arbitrary
long times.
The present model serves two purposes: On one hand, it provides a concrete route
where to search for signatures of entanglement in biomolecular
systems operating at room temperature. On the other hand,
it indicates new directions to design molecular scale  machines able
to generate non-trivial quantum states even in noisy environments.
Finally, let us remark that an experimental simulation of our model
is achievable with state-of-the-art technology in setups like ion traps
or nano-mechanical systems.

\section*{Acknowledgements}

This work was supported by the FWF (SFB FoQuS, J.M.C. through the
Lise Meitner Program and a Marie-Curie Intra-European Fellowship
within the 7th European Community Framework Programme);
S.P. acknowledges support from the UK EPSRC QIPIRC grant.


\section*{References}

\begin{thebibliography}{99}

\bibitem{Jang04} Jang S, Newton M D and Silbey R J 2004 Phys. Rev. Lett. \textbf{92} 218301

\bibitem{Fleming04} Fleming G R and Scholes G D 2004 Nature \textbf{431} 256

\bibitem{Fleming07} Engel G S, Calhoun T R, Read E L, Ahn T K, Mancal T, Cheng Y C, Blankenship R E and Fleming G R
2007 Nature \textbf{446} 782

\bibitem{Lee07} Lee H, Cheng Y C and Fleming G R 2007 Science \textbf{316} 1462

\bibitem{Scho09} Collini E and Scholes G D 2009 Science \textbf{323} 369

\bibitem{Eng10} Panitchayangkoon G, Hayes D, Fransted K A, Caram J R, Harel E, Wen J, Blankenship R E and Engel G S
2010 Proc. Natl. Acad. Sci. USA \textbf{107} 12766

\bibitem{Scho10} Collini E, Wong C Y, Wilk K E, Curmi P M G, Brumer P and Scholes G D 2010 Nature \textbf{463} 644

\bibitem{Mohseni0805} Mohseni M, Rebentrost P, Lloyd S and Aspuru-Guzik A 2008 J. Chem. Phys. \textbf{129} 174106

\bibitem{Rebentrost08} Rebentrost P, Mohseni M and Aspuru-Guzik A 2009 J. Phys. Chem. B \textbf{113} 9942;
Rebentrost P, Mohseni M, Kassal I, Lloyd S and Aspuru-Guzik A 2009 New J. Phys. \textbf{11} 033003

\bibitem{Plenio0807} Plenio M B and Huelga S F 2008 New J. Phys. \textbf{10} 113019

\bibitem{Olaya} Olaya-Castro A, Lee C F, Fassioli Olsen F and Johnson N F 2008 Phys. Rev. B \textbf{78} 085115

\bibitem{Buchleitner} Scholak T, de Melo F, Wellens T, Mintert F and Buchleitner A 2011 Phys. Rev. E \textbf{83} 021912

\bibitem{Eisfeld} Briggs J S and Eisfeld A 2011 Phys. Rev. E \textbf{83} 051911

\bibitem{Ste89} Steiner U E and Ulrich T 1989 Chem. Rev. \textbf{89} 51

\bibitem{Schu} Schulten K, Swenberg C E and Weller A 1978 Z. Phys. Chem \textbf{NF111} 1

\bibitem{Ritz} Ritz T, Adem S and Schulten K 2000 Biophys. J. \textbf{78} 707

\bibitem{Hore} Rodgers C T and Hore P J 2009 Proc. Natl. Acad. Sci. USA \textbf{106} 353

\bibitem{GG10} Cai J M, Guerreschi G G and Briegel H J 2010 Phys. Rev. Lett. \textbf{104} 220502

\bibitem{Vedral11} Gauger E M, Rieper E, Morton J J L, Benjamin S C and Vedral V 2011 Phys. Rev. Lett. \textbf{106} 040503

\bibitem{Hans0806} Briegel H J and Popescu S 2008 arXiv:0806.4552

\bibitem{Cai10} Cai J M, Popescu S and Briegel H J 2010 Phys. Rev. E \textbf{82} 021921

\bibitem{Alberts08} Alberts B, Johnson A, Lewis J, Raff M, Roberts K and Walter P 2008
\emph{Molecular Biology of the Cell}, (New York: Garland Science)

\bibitem{Wootters98} Wootters W K 1998 Phys. Rev. Lett. \textbf{80} 2245

\bibitem{Tong2005} Tong D M, Singh K, Kwek L C and Oh C H 2005 Phys. Rev. Lett. \textbf{95} 110407

\bibitem{Tong2007} Tong D M, Singh K, Kwek L C and Oh C H 2007 Phys. Rev. Lett. \textbf{98} 150402

\bibitem{BreuerBook} Breuer H P and Petruccione F 2002 \textit{The Theory of Open Quantum Systems} (New York: Oxford Univerity Press)

\bibitem{Hans09} Briegel H J and Popescu S 2009 arXiv:0912.2365

\bibitem{FeynmanBook} Feynman R P and Hibbs A R 1965 \textit{Quantum Mechanics and Path Integrals} (New York: McGraw-Hill)

\bibitem{Feynman63} Feynman R P and Vernon F L 1963 Ann. Phys. \textbf{24} 118

\bibitem{Makri} Makarov D E and Makri N 1994 Chem. Phys. Lett. \textbf{221} 482;
Makri N and Makarov D E 1995 J. Chem. Phys. \textbf{102} 4600;
Makri N and Makarov D E 1995 J. Chem. Phys. \textbf{102} 4611

\bibitem{Thorwart11} Eckel J, Reina J H and Thorwart M 2011 New J. Phys. \textbf{11} 085001

\bibitem{Thorwart06} Eckel J, Weiss S and Thorwart M 2006 Eur. Phys. J. B \textbf{53} 91

\bibitem{Cals05} Calsamiglia J, Hartmann L, D\"{u}r W and Briegel H J 2005 Phys. Rev. Lett. \textbf{95} 180502

\bibitem{Lor06} Hartmann L, D\"{u}r W and Briegel H J 2006 Phys. Rev. A. \textbf{74} 052304

\bibitem{Lor07} Hartmann L, ~D\"{u}r W and Briegel H J 2007 New. J. Phys. \textbf{9} 230

\bibitem{Hans93} Briegel H J and Englert B G 1993 Phys. Rev. A \textbf{47} 3311

\bibitem{Cals07} Calsamiglia J, Hartmann L, D\"{u}r W and Briegel H J 2007 Int. J. Quant. Inf. \textbf{5} 509

\end{thebibliography}
\end{document}